\documentclass[letterpaper,11pt]{article}

\usepackage{epsfig, amsmath, amssymb,array}

\begin{document} 

\baselineskip=18pt

%%%%%%%%%%
%%%%%%%%%%    Title page
%%%%%%%%%%

\thispagestyle{empty}
\vspace{20pt}
\font\cmss=cmss10 \font\cmsss=cmss10 at 7pt

\begin{flushright}
\today \\
UMD-PP-09-030\\
\end{flushright}

\hfill
\vspace{20pt}

\begin{center}
{\Large \textbf
{Relaxing Constraints from 
%
%(Charged) 
%
Lepton Flavor Violation
in 
$5D$ Flavorful Theories
%
%Warped Extra Dimension
%
}
}
\end{center}

\vspace{15pt}

\begin{center}
{\large Kaustubh Agashe\footnote{email: kagashe@umd.edu}
%
%$\, ^{}$
%
} \\
\vspace{15pt}
%
%$^{}$
%
\textit{Maryland Center for Fundamental Physics,
     Department of Physics,
     University of Maryland,
     College Park, MD 20742, USA}

\vskip .3cm
%
%\centerline{\tt kagashe@umd.edu}
%

\end{center}

\vspace{20pt}

\begin{center}
\textbf{Abstract}
\end{center}
\vspace{5pt} {\small \noindent 
%
%The framework of a warped extra dimension provides solutions to both 
%the Planck-weak and 
%flavor hierarchies of the Standard Model (SM), the latter via
%profiles for the SM fermions in the extra dimension.
%
%(or Extra dimension provides solution to
%flavor hierarchy via profiles...)
%
%Such a scenario results in flavor violation from the
%non-universal couplings of SM fermions to 
%
%gauge 
%
%the Kaluza-Klein
%(KK) modes of the SM gauge and fermion fields.
%
%We discuss
%how new choices of representations under the
%extended $5D$ electroweak gauge symmetry $\Big[ SU(2)_L \times
%SU(2)_R \times U(1)_X \Big]$ (required to satisfy electroweak precision tests)
%for the SM fermions
%can ameliorate constraints on
%KK mass scale 
%from charged lepton flavor violation,
%especially in the models accounting for neutrino masses
%and mixings.
%%
%%(or from TeV-scale new states...)
%%
%Such non-canonical/non or less-minimal representations
%allow small mxing angles for left-handed (LH) charged leptons, while
%simultaneously mixing angles for their $SU(2)_L$ partners, i.e.,
%the LH neutrinos can be large. 
%
We 
propose
%
%discuss 
%
new mechanisms for 
ameliorating the constraints on
the 
Kaluza-Klein 
(KK) mass scale 
from charged lepton flavor violation
in the framework of the Standard Model (SM) fields
propagating in a warped
extra dimension,
especially in models accounting for neutrino 
data.
%
%masses
%and mixings. 
%
These mechanisms utilize
%
%rely on 
%
the 
extended
five-dimensional ($5D$) electroweak gauge symmetry $\Big[ SU(2)_L \times
SU(2)_R \times U(1)_X \Big]$ 
%
%which might already be required
%
which is already strongly
motivated in order to satisfy electroweak precision tests
in this framework.
We show that new choices of 
%
%non-canonical/non or less-minimal 
%
representations for leptons 
under this symmetry 
(naturally) can allow small mixing angles for left-handed (LH) charged leptons 
and {\em simultaneously} large mixing angles for their $SU(2)_L$ partners, i.e.,
the LH neutrinos, with the neutrino data being 
%
%explained
%
accounted for by the latter
mixings.  
%
%preventing 
%
Enhancement of charged lepton flavor violation by the 
large mixing angle observed
in
leptonic charged currents, which is
present 
%
%in other models 
%
for 
%
%canonical
%
the minimal
%
%simplest 
%
choice of representations
where the 
%
%two 
%
LH charged lepton and neutrino 
mixing angles are similar,
can thus be avoided in these models. 
This idea might also be useful for
suppressing 
%
%some/sub-dominant contributions to
%flavor violation in the quark sector 
%
the contributions to $B_{ d, \; s }$ mixing in this framework
and in order to suppress
flavor violation 
%
%(especially in charged lepton sector)
%
from exchange of superpartners (instead of from KK modes) in $5D$ ``flavorful 
supersymmetry''
%
%SUSY 
%
models.
%
%This idea can also be applied to
%
%other
%extra-dimensional models which account for flavor hierarchy via profiles
%in order to 
%suppress flavor violation,
%for example, flavorful SUSY, where the dominant flavor violating effects
%involve superpartners (instead of KK modes).
%
Additionally,
%
%independently, 
%
the less minimal representations
can provide custodial protection for shifts in couplings of
fermions to $Z$
and, in turn, further suppress charged lepton flavor-violation from tree-level $Z$
exchange in the warped extra-dimensional scenario.
As a result, 
%
%a few 
$\sim O(3)$
TeV KK mass scale can be 
%
%easily} 
%
%allowed
%by 
%
simultaneously
consistent with charged lepton flavor violation
%
%consistent 
%
%with 
%
{\em and} neutrino data, even without any 
%
%flavor symmetries/
%
particular
structure
in the $5D$ flavor parameters in the framework of a warped extra dimension.
}

\vfill\eject
\noindent

%%%%%%%%%%
%%%%%%%%%%    Main Text
%%%%%%%%%%

%%%%%%%%%%%%%%%%%%%%%%%%%%%%%%%%%%%%%%%%%%%%%%%%%%%%%%%%%%%%%%%%%%%%%%%%%%%%%%%%%
\section{Introduction}

The framework of a warped extra dimension 
was proposed in order to provide a solution to the 
Planck-weak hierarchy problem of 
the Standard Model (SM) 
\cite{Randall:1999ee}.
With the SM
gauge and fermions fields propagating in the extra dimension 
\cite{bulk, gn, gp}, it 
can also account for the flavor hierarchy
of the SM via extra-dimensional profiles
for SM fermions \cite{gn, gp}. Inherent to this approach is 
flavor violation from the resulting non-universal
couplings of SM fermions to the Kaluza-Klein (KK) modes
of the SM fields which are the new particles
present in this framework \cite{Delgado:1999sv}.
In spite of an
analog of Glashow-Iliopoulos-Maiani (GIM) mechanism being 
%
%inherent
%
built-in to
this framework \cite{gp, Huber:2000ie, Agashe:2004cp}, the lower limits 
on the mass scale of the KK gauge bosons
from the flavor violation in the {\em quark} sector can still be 
$\sim O(5-10)$ TeV\footnote{A clarification
about notation is in order here. The uncertainty
in the bounds on KK mass scale from flavor violation  
denoted by the symbol ``$\sim...$''
comes from effects of modifications to the minimal model
such as
brane-localized kinetic terms for bulk fields 
\cite{Davoudiasl:2002ua} or replacement of the endpoint of the extra dimension
by a ``soft wall'' \cite{soft}. Such variations 
are present even for limits on KK mass scale
from electroweak precision tests. On the other hand, the
symbol
``$O(...)$'' refers to uncertainties in the bounds from flavor violation
due to presence of 
$O(1)$ factors in $5D$ Yukawa (which is an inherent feature of
$5D$ flavor ``anarchy'') and due to the presence (typically) of more than one 
term (of similar size, but uncorelated) in the flavor-violating amplitude.
Contributions to electroweak precision
tests are not very sensitive to the latter
types of effects and hence the
``$O(...)$'' factor is absent in that case.}, 
depending on the details of the model
\cite{Csaki:2008zd, Blanke:2008zb, Agashe:2008uz}
(see also \cite{Fitzpatrick:2007sa, Davidson:2007si}). 
Whereas, 
flavor violation in the charged lepton sector requires
a gauge KK mass scale at least as large as $\sim O(5)$ TeV {\em without
consideration of neutrino data} \cite{Agashe:2006iy}.
It turns out that, in {\em minimal} models,
enhancement of charged lepton flavor violation
by the large mixing required in order to account
for neutrino data
results in constraints
being even {\em stronger} than $\sim O(5)$ TeV \cite{Perez:2008ee}.

Such a large gauge KK scale might imply a tension with a
%
%{\em natural 
%
resolution of the Planck-weak
hierarchy problem of the SM which requires a KK scale $\sim$ TeV.
Also, signals from direct production of
these KK modes at the Large Hadron Collider (LHC) (including upgrades) 
are then extremely challenging (if not unlikely).
Therefore, it is very interesting to study mechanisms to ameliorate constraints
from this flavor violation, thus
allowing for a
lower gauge KK mass scale.
Recently, five-dimensional ($5D$) flavor symmetries (for both
quark and lepton sectors) have been suggested
for this purpose \cite{Cacciapaglia:2007fw, 
Fitzpatrick:2007sa, 
Chen:2008qg, Perez:2008ee, Csaki:2008eh}
such that a gauge KK scale as low as $\sim O(3)$ TeV might be allowed.

In this paper, we propose {\em alternative}
%
%new 
%
mechanisms
in order to suppress charged {\em lepton} flavor violation.
The idea is to 
use new (less minimal) representations 
for leptons under the extended electroweak (EW) $5D$ gauge
symmetry\footnote{Very recently in reference \cite{Carena:2009yt}, such non-minimal
representations for leptons were studied in the context of 
gauge-Higgs unification, 
but their relevance for suppression of charged 
lepton flavor violation was not discussed.} -- such an extended symmetry is 
{\em in any case} strongly motivated 
in order to suppress 
contributions to electroweak precision tests, in
particular, the $T$ parameter \cite{Agashe:2003zs}. We demonstrate that 
\begin{itemize}

\item

certain such choices of representations
can allow small and large
mixing angles to {\em naturally} co-exist 
for left-handed (LH) charged leptons
and neutrinos, respectively, in spite
of them being $SU(2)_L$ partners.

\end{itemize}
Thus 
it is possible (unlike in the minimal models) to avoid the large mixing angles required
to explain the neutrino data from 
%
%feeding into
%
exacerbating 
charged lepton flavor
violation.\footnote{Recently \cite{Agashe:2008fe}, 
it was shown that with a profile for the SM Higgs in the extra dimension
(but still peaked near the endpoint of the extra dimension) 
\cite{Davoudiasl:2005uu}
(instead of a $\delta$-function
localized Higgs) and with SM neutrinos being Dirac, it is possible to 
achieve a similar ``decoupling''
of LH neutrino and charged lepton mixing angles.}
As a result, even after
including neutrino data, the constraint on the gauge KK scale
can relax to the $\sim O(5)$ TeV value in the case without neutrino data.
This new idea
can suppress 
contributions to $B_{ d, \; s }$ mixing (which, however, are not the dominant
constraints from flavor violation in quark sector)
%
%some/certain/sub-dominant contributions to 
%quark flavor violation 
%
as well by allowing 
LH down-type and up-type quark mixing angles to 
be parametrically different.
Similarly it can also be applied to other extra-dimensional models
(which account for flavor hierarchy via profiles for SM fermions
in the extra dimension)
in order to 
suppress flavor violation (especially in charged lepton sector).

{\em Independently} of the decoupling
of large neutrino mixings from charged lepton
sector, we show that 
\begin{itemize}

\item
such new representations 
can result in a custodial
symmetry which protects
shifts in coupling of SM fermions to $Z$ in
the framework of a warped extra dimension \cite{Agashe:2006at}.

\end{itemize}
Hence flavor-violating $Z$ couplings to leptons 
and the resulting tree-level $Z$ 
exchange contributions to processes such as
$\mu$ to $e$ conversion in nuclei and $\mu \rightarrow
3 \; e$ can be suppressed\footnote{The role of 
such a custodial symmetry
in suppressing flavor violation in the {\em quark} sector
was discussed in \cite{Blanke:2008zb, Blanke:2008yr}.}.
{\em Combining} the above two ideas, we show that it is 
possible to 
reduce the lower limit on gauge KK scale from charged lepton
flavor violation (including neutrino data) down to
$\sim O(3)$ TeV
from 
$>O(5)$ TeV in the minimal model.
%
%$\sim O(5)$ TeV scale without neutrino masses.
%

The outline of the rest of the paper is as follows.
We begin with an overview of the framework
of warped extra dimension (including both the
quark and lepton sectors) and a {\em qualitative} outline 
of the problem of charged lepton flavor violation and the solutions
proposed in this paper.
Then we
present 
%
%numerical
%
{\em quantitative} estimates
for charged lepton flavor violation in section \ref{estimate}, including how 
charged lepton flavor violation is enhanced by
large neutrino mixing. The
{\em central observations} of this paper are in the next two sections: 
in section \ref{decoupling}, we show how to decouple mixings
of the LH charged lepton and neutrino sectors, with many example representations
for leptons under the extended $5D$ 
EW
%
%electroweak 
%
gauge symmetry: the general
idea is illustrated in Fig. \ref{trick}. 
In section \ref{tuning}
we consider a choice of profiles in order to 
obtain large neutrino mixing with mild tuning and then
discuss the custodial protection mechanism which is critical to suppressing 
flavor-violating couplings to $Z$
with this choice of profiles. 
A summary of various possibilities along these lines is provided 
in table \ref{choice}.
We conclude in section \ref{concl}
with a brief discussion
%
%mention 
%
of signals
for our new models. We also comment on 
the applicability of the mechanisms presented here to
suppressing some contributions to 
flavor violation in the quark sector
and to other 
extra-dimensional scenarios such as $5D$ ``flavorful supersymmetry'' (SUSY).

\section{Overview}

A
slice of anti-de Sitter (AdS) space in $5D$ \cite{Randall:1999ee} provides
a solution to the Planck-weak
hierarchy problem of the SM. 
%
%In short/
%
Basically, the warped
geometry implies that the UV cut-off of the effective $4D$ theory
depends on location in the extra dimension ($y$): $M_{ 4D \; eff. }
\sim M_{ 5D } e^{ - k y }$, where $k$ is the AdS curvature scale, $e^{ - k y }$
is called the warp factor
and $M_{ 5D }$ is the fundamental $5D$ mass scale.
The $4D$ graviton (zero-mode of the $5D$ gravitational field) is automatically
localized near the $y=0$ end of the extra dimension
(hence called the Planck/UV brane). Suppose the Higgs 
sector is taken to be localized near the
other end of the extra dimension (called the TeV/IR brane):
$y = \pi R$, where $R$ is the proper 
size (or radius) of the extra dimension -- such a
localization happens automatically if
Higgs is the $5^{th}$ component of a 
$5D$ gauge field \cite{Contino:2003ve}. Then
the Planck-weak hierarchy can be explained by a mild hierarchy between
the AdS curvature radius ($\sim 1 / k$ which is taken to be 
$\sim 1 / M_{ 5D }$)  and $R$:
$M_{ Higgs } \; \hbox{or} \; M_{  weak }
\sim M_{ 5D } e^{ - k \pi R }$. Note that $M_{ 5D }
\sim M_{ Pl } \sim 10^{ 18 }$ GeV is required in order to
reproduce the observed ($4D$) Planck scale due to warp factor being $1$
at the location of the $4D$ graviton.
In turn, such a mild hierarchy between the proper size 
of the extra dimension and curvature radius:
$ k \pi R \sim \log \left( M_{ Pl } / \hbox{TeV} \right)
\sim 30$
can be stabilized by suitable 
mechanisms \cite{Goldberger:1999uk}.
It is also
that interesting that, based on the AdS/CFT correspondence \cite{Maldacena:1997re}, such a 
scenario is conjectured to be dual to SM Higgs being a 
composite of TeV-scale strong dynamics 
\cite{Arkani-Hamed:2000ds, Contino:2003ve}.

\subsection{SM in the bulk of warped extra dimension}

Such a framework can also provide a 
solution to the flavor hierarchy of the SM if the SM fermions
arise as zero-modes of fermions propagating in the extra dimension
\cite{gn, gp}. Namely, the
profiles of the SM fermions in the extra dimension 
are then controlled by their $5D$ masses.
The crucial feature is that 
small variations in the $5D$ masses enable the
SM fermions to have profiles which are peaked either near the 
Planck or TeV branes or are flat. 
This feature results in small/large/intermediate couplings
of the SM fermions
to the SM Higgs (which is 
localized near the TeV brane), simply based on overlaps of profiles
in the extra dimension, 
i.e., without any hierarchy in the fundamental ($5D$) 
parameters (Yukawa couplings and $5D$ fermion masses).

SM gauge fields must also then originate as zero-modes of $5D$ fields 
(``SM in the bulk'') \cite{bulk, gp} --
it turns out that the 
SM gauge fields have a flat profile in the extra dimension.
In addition to the zero-modes, the $5D$ fields have
other, non-trivial excitations in
the extra dimension (called Kaluza-Klein or KK modes) which
appear as heavier particles from the $4D$ point of view.
In the warped case, these KK modes 
turn out to be automatically localized near the TeV brane
and have masses $\sim k e^{ - k \pi R }$, i.e., at the $\sim$ TeV-scale.
Thus all SM particles (except perhaps the SM Higgs)
have KK modes in this scenario.
So, contributions
from these KK modes to precision tests of the SM
can constrain this scenario.
In particular, 
electroweak precision tests (EWPT)
can be under control, using custodial symmetries
to protect contributions to the
$T$ parameter \cite{Agashe:2003zs}
and the
shift in $Z b \bar{b}$ coupling \cite{Agashe:2006at}, even with KK masses
of a few (or several) TeV \cite{Agashe:2003zs, EWPTmodel, Contino:2006qr}.

\subsection{Flavor violation}

More relevant to this paper, there is 
flavor violation from exchange of
KK modes which necessarily have
non-universal couplings to the SM fermions (given that the flavor hierarchy is accounted
for by SM fermions' non-universal profiles)
\cite{Delgado:1999sv}.
However, there is an
%
%built-in
%
analog of the GIM mechanism of
the SM which is automatic in this scenario since the
non-universalities in the couplings of SM fermions to
KK modes are of size of $4D$ Yukawa couplings (due to
KK's having similar profile to Higgs) \cite{gp, Huber:2000ie, Agashe:2004cp}. 
However, 
even in the presence of this RS-GIM mechanism, 
recently 
\cite{Csaki:2008zd, Blanke:2008zb} (see also
\cite{Fitzpatrick:2007sa, Davidson:2007si})
it was shown that the constraint on the
KK mass scale from tree-level contributions of KK {\em gluon} 
to $\epsilon_K$ is
quite stringent.
In particular, for the model with the
SM Higgs (strictly) localized on the TeV brane, 
%
%in a $5D$ slice
%of anti-de Sitter space (AdS),
%
the limit on the KK gluon 
mass scale from $\epsilon_K$ is $\sim O(10)$ TeV
%
%(constraints from rest
%of the $\Delta F =2$ processes are then satisfied)
%
for the smallest allowed $5D$ QCD coupling
obtained by {\em loop}-level matching to the $4D$ coupling with negligible
tree-level brane kinetic terms. 
%
%(in the framework which addresses the
%Planck-weak hierarchy).
%
On the other hand, for larger brane kinetic terms such
that the $5D$ QCD coupling (in units of
the AdS curvature scale, $k$) is
$\sim 4 \pi$, the lower limit on KK gluon mass scale increases to $\sim O(40)$ 
TeV.
In addition, the constraint on the KK gluon mass scale
is weakened as the size of the $5D$ Yukawa (in units of 
$k$)
%
%the
%AdS curvature scale) 
%
is increased.
However,
this direction reduces the regime of validity of the
$5D$ effective field theory (EFT):
the above limits on KK gluon mass scale are for the size of $5D$ Yukawa 
such that about two KK modes are allowed in the $5D$ EFT.

Whereas, with a profile for the SM Higgs in the extra dimension (but 
which is still peaked near
TeV brane \cite{Davoudiasl:2005uu}, called a ``bulk Higgs'') and choosing
the smallest allowed $5D$ QCD coupling
and two KK modes in the $5D$ EFT, it was demonstrated in reference
\cite{Agashe:2008uz} that
$\sim O(3)$ TeV KK gluon mass scale can be consistent with
$\epsilon_K$\footnote{It was also argued in the same reference
that a larger size of the $5D$ QCD coupling might in fact
conflict with $5D$ perturbativity.}.
However, in the ``two-site'' model \cite{Contino:2006nn}
(which is 
an economical approach to studying 
this
%
%the
%$5D$ AdS 
%
framework 
by restricting to the SM
fields and their first KK excitations), it was also shown in
reference \cite{Agashe:2008uz} that there is a ``tension''
between satisfying constraints form $\epsilon_K$ and
BR$\left( b \rightarrow s 
\gamma \right)$ (the latter observable being sensitive to loop
effects of KK modes). Thus, the limit on the mass scale for the new particles
(assuming the heavy fermions and gauge bosons have same mass)
must actually be a bit larger, namely, $\sim O(5)$ TeV
to be consistent with this {\em combination} of constraints.
Hence, it was also
suggested reference \cite{Agashe:2008uz} that the $5D$  
models with a bulk Higgs can allow a similar, i.e., $\sim O(5)$ TeV, 
gauge KK scale
to be consistent with the entire body of data on flavor violation
in the quark sector.

Furthermore, $5D$ 
flavor symmetries 
%
%can ameliorate
%these constraints from flavor violation 
%
in the
quark sector can add more
structure to the $5D$ model, for example, by relating
the $5D$ (or bulk) fermion masses to the $5D$ Yukawa couplings and/or
by enforcing 
degenerate bulk masses \cite{Cacciapaglia:2007fw, Fitzpatrick:2007sa,
Csaki:2008eh}. 
Such a reduction in the number of flavor parameters
results in suppressed quark sector flavor violation.
Also, by lowering the UV-IR hierarchy, i.e.,
$k \pi R$, it is possible to
lower the gauge KK scale allowed by quark sector flavor violation 
\cite{Davoudiasl:2008hx}, although in
this paper we will always assume Planck-weak hierarchy, 
i.e., $k \pi R \sim 30$.
Further studies
of flavor violation, especially experimental signals, 
appear in references 
\cite{others1, others2}.

In this paper, we focus instead on 
flavor violation and hierarchy of masses in the
{\em lepton} sector. With{\em out} 
consideration of neutrino data, it was shown in
reference \cite{Agashe:2006iy} that the constraint from charged lepton 
flavor violation on gauge KK mass scale is 
%
%several, i.e.,
%
$\sim O(5)$ TeV -- such a 
strong constraint is mainly due to a tension
between the two processes $\mu$ to $e$ conversion in nuclei 
(which occurs at tree-level in this framework) and loop-induced
$\mu \rightarrow e \gamma$. 
Note that this constraint was obtained for 
hierarchies in charged lepton
masses being explained by the choice of hierarchical profiles near the
TeV brane for both
right-handed (RH) and LH charged leptons
so that both RH and LH charged lepton mixing angles (given by ratio of 
respective profiles at the TeV brane) were set to be small (roughly
{\em square root} of ratio of charged lepton masses).

\subsection{Charged lepton flavor violation and neutrino data}
\label{exacerbate}

However, including neutrino data, two new and distinct issues 
come up (see also related discussion in reference \cite{Perez:2008ee}):
\begin{itemize}
\item[(i)] {\bf 
%
%Effect of 
%
Enhancement due to 
large mixing angle}: With the 
%
%canonical
%
simplest representations under the extended bulk EW gauge
symmetry, i.e.,
$SU(2)_L \times SU(2)_R \times U(1)_X$ (such an extension is
typically required to satisfy EWPT) and a (strictly) TeV brane-localized
Higgs, the 
charged lepton and neutrino
(Dirac) masses originate from the {\em same} LH lepton bulk profiles evaluated
at the
TeV brane.
Thus
the mixing angles for LH charged leptons and neutrino are similar and, in turn, 
a combination these two mixing angles is what enters
charged current lepton interactions.
So, this mixing is required to be large 
in order to explain neutrino oscillation data.

Such large LH
charged lepton mixing results in 
an enhancement of 
charged lepton flavor violation {\em relative to without
considerations of neutrino data} as in \cite{Agashe:2006iy} --
as mentioned above, in reference \cite{Agashe:2006iy}
{\em both} RH and LH
charged lepton mixing
was set to be small. Thus, the gauge
KK scale is constrained to be {\em larger} than 
%
%several 
%
$\sim O(5)$
TeV 
%
%is required 
%
in order 
to be consistent
with all the data, i.e., 
charged lepton flavor violation 
%
%once we include 
%
{\em and}
neutrino mixings.
\item[(ii)] {\bf Flat profiles for mild tuning}:
For the case of a brane-localized Higgs,
large LH neutrino mixing clearly requires non-hierarchical
(i.e., with $\sim O(1)$ ratios)
profiles for LH leptons
near the TeV brane where the $4D$ Yukawa coupling originates.
However, 
if the LH lepton profiles are peaked near the Planck brane, i.e., 
{\em exponentially} suppressed near the TeV brane, then
it is clear that we need to tune 
the bulk masses (which control the
exponentials) to be (almost) degenerate in order for the profiles near 
the TeV brane
to be non-hierarchical.

{\em If we require no tuning of bulk masses}, then we might be forced to
choose 
close-to-flat profiles for all generation LH leptons
such that 
the profiles near the TeV brane
can be non-hierarchical with only a {\em mild} tuning of bulk masses.
However, such a choice results in a
larger coupling of SM leptons to gauge KK modes 
(which are localized near IR brane) relative to 
the case of without
considerations of neutrino data, i.e., where lepton profiles -- both LH and RH -- 
are peaked near the Planck brane (motivated by smallness of charged lepton masses).
In turn, the larger couplings of leptons to KK modes
enhance charged lepton flavor violation
via tree-level $Z$ exchange further, i.e.,
{\em in addition} to the effect of large LH charged lepton mixing angles
mentioned in point (i) above.

\end{itemize}

Invoking $5D$ flavor symmetries is one way
to 
%
%ameliorate/
%
solve 
the above problems \cite{Chen:2008qg, Perez:2008ee}. 
In particular,
even if LH lepton profiles are 
peaked near the Planck brane, the 
(almost) degenerate LH lepton bulk masses required to give
non-hierarchical profiles near TeV brane (and hence
large mixing) are then enforced by a symmetry.
Also, the resulting {\em universal} couplings of
LH charged leptons to gauge KK modes (GIM
mechanism) suppress LH charged lepton 
flavor violation from zero-KK gauge boson
mixing: see top right-hand side of Fig. \ref{diagram}.
Independently, such symmetries can relate
bulk masses to $5D$ Yukawa couplings
(just like for quarks discussed above) thus 
reducing the number of flavor parameters
(i.e., adding
structure). Hence 
%
%further 
%
LH charged lepton flavor violation from zero-KK {\em fermion}
mixing (see top left-hand side of Fig. \ref{diagram}) 
and similarly $\mu \rightarrow e \gamma$ are suppressed
as well.

Alternatively \cite{Agashe:2008fe}, for Higgs with a profile in
the bulk (but still peaked near the TeV brane)
\cite{Davoudiasl:2005uu} and with
neutrinos being Dirac particles, neutrino masses of the 
observed size 
(i.e., 
required
to account for neutrino oscillations) 
%
%sufficiently large 
%
can
arise from overlap near
the {\em Planck} 
brane, whereas charged lepton masses originate (as usual) from the
overlap of profiles near the TeV brane. Then the
much smaller neutrino masses (relative to charged lepton) and
large vs. small mixing in neutrino and charged lepton sectors arise naturally.
The point is that the LH lepton profiles can be non-hierarchical and large near the
Planck brane
(giving large mixing for LH neutrinos and 
ultra-small masses due to small Higgs profile at the Planck brane), while simultaneously
being  
small and hierarchical
near the TeV brane (giving small mixing and small masses for LH charged leptons).

\subsection{New $5D$ gauge representations for leptons}

In this paper, we propose an alternative to both the above ideas
to suppress charged lepton flavor violation while
obtaining large neutrino mixings.
We  
still consider neutrino masses originating from near the {\em TeV} 
brane (say, Higgs
is localized 
on the TeV brane or it leaks into the bulk, but not 
sufficiently for Dirac neutrino masses
from overlap near the Planck brane to be 
%
%significant
%
of the observed size). 
The new idea is to 
use 
%
%non or 
%
less minimal
%
%non-canonical 
%
representations
under the $SU(2)_R \times U(1)_X$ gauge symmetry.
In particular, there are two new mechanisms as follows.
\begin{itemize}
\item[(a)] {\bf Decoupling large neutrino mixing from
charged lepton masses}: the idea is that
LH lepton zero-mode for each generation can arise
as a {\em combination} of zero-modes
from 2 different $5D$ multiplets: see Fig. \ref{trick}.
Such a scenario allows LH mixing angles to be parametrically different
for charged leptons vs. neutrinos
since the two mass matrices (and hence mixing angles) can originate from
the two different components of the LH lepton zero-mode. 
In particular, mixing angles can then
be 
small for charged vs. large for neutrinos.
This novel possibility
prevents large mixing angles in leptonic charged
currents from infiltrating both tree-level $Z$ exchange
(giving $\mu$ to $e$ conversion in nuclei) and 
loop-induced dipole operators
(giving $\mu \rightarrow e \gamma$).
\item[(b)] {\bf Custodial protection}:
Independently, some choices of representations under the
extended bulk EW gauge symmetry
can result in a custodial symmetry for the shift in the couplings of 
leptons
%
%fermions
%
to $Z$
(similar to one used to suppress shift in $Z b \bar{b}$ 
\cite{Agashe:2006at}). Such 
a symmetry 
can then suppress the flavor-violating couplings of leptons to
$Z$ and hence charged lepton flavor violation via tree-level
$Z$ exchange. Such a suppression is especially desirable if we choose 
(close-to-) flat profiles in order to generate large neutrino mixings without
tuning (as mentioned
in point (ii) in section \ref{exacerbate}). In such a case, the enhanced
coupling of charged leptons to KK modes is 
still problematic for charged lepton flavor
violation (as discussed in point (ii) in section \ref{exacerbate}), even if we obtain {\em small} charged lepton mixing angles
using the idea in point (a) above.

\end{itemize}
It is in fact possible to {\em combine} the above two features
for some choice of representations of leptons
under the $5D$ 
%
%electroweak 
%
EW
%
%$SU(2)_R \times U(1)_X$ 
%
gauge symmetry, resulting in
$\sim O(3)$
%
%{\em a few} 
%
TeV KK scale being consistent with
charged lepton flavor violation and large neutrino mixings (without
any particular structure in the $5D$ flavor parameters).
Various cases utilizing the above two ideas: (a) and/or (b) 
are listed in table \ref{choice}.

\section{Estimates for charged lepton flavor violation}
\label{estimate}

%-- summary of framework of warped extra dimension (refer to for more
%details):
%SM gauge and fermion fields as
%zero-mode of fields in bulk, Higgs localized on/near IR brane
%
%-- SM gauge fields have flat profile, SM fermion profile
%controlled by bulk mass -- small variations result
%in peaker near IR/UV brane, giving small/large coupling
%to Higgs)...
%
%-- all (light) KK's peaked near IR brane (like Higgs)
%
%-- flavor-violation due to non-universal couplings to KK's, but
%non-universality $\propto$ SM Yukawa giving analog of GIM...
%

In this section, we collect formulae 
for charged lepton flavor violation valid for the general case and then
specialize to the
models with neutrino masses.
Since we are mainly concerned with
parametric effects and mechanisms, estimates of these
effects (i.e., formulae valid up to $O(1)$ factors) will suffice
for our purpose.
For more detailed formulae, the reader is referred to
previous studies (see references \cite{Agashe:2004cp, Agashe:2006iy}
for example).
In addition to the effects of KK modes summarized below,
there are also operators induced by physics at the cut-off
of the $5D$ theory.
For simplicity, we assume here that
we have a bulk Higgs (but with a profile which is peaked
near the TeV brane), where such cut-off effects can
be shown to be smaller than KK-induced ones
(see references \cite{Agashe:2004cp, Agashe:2006iy, Perez:2008ee}).

We first perform a KK decomposition for SM gauge and fermion
fields setting the Higgs vev to zero.
The $4D$ Yukawa coupling, i.e., the coupling
of SM Higgs to two zero-mode fermions (say charged leptons), is given by:
\begin{eqnarray}
Y_4 \left( c_{ e_L },  c_{ e_R } \right)
& \sim Y_5 \sqrt{k} f \left( c_{ e_L } \right) f \left( c_{ e_R } \right)
\label{y4d}
\end{eqnarray}
where 
$Y_5$ is $5D$ Yukawa coupling of mass dimension $-1/2$\footnote{due
to SM Higgs being in the bulk.} and
$f$'s are
ratio of zero-mode and
KK profiles near the TeV brane:
\begin{eqnarray}
f (c) & \approx & \left\{ 
\begin{array}{c} 
\sqrt{
\left( c - \frac{1}{2}
\right) e^{ k \pi R \left( 1 - 2 c \right) }              
}
%
%e^{ - k \pi R \left( c - 1/2 \right) } 
%
\; \hbox{for} 
\; c > 1/2
               \\
\sqrt{
\frac{1}{2 k \pi R }
}
%
%              \frac{1}{ \sqrt{ k \pi R } } 
%
\; \hbox{for} \; c = 1/2 \\
\sqrt{
\left( \frac{1}{2} - c \right)
}
\; \hbox{for} \; c < 1/2
    \end{array} 
\right.
\label{f}
\end{eqnarray}
where $c$ is the $5D$ mass for the corresponding
$5D$ fermion in units of $k$.
We can show that the KK Yukawa coupling, i.e., the coupling of Higgs
to two KK fermions, is given by:
\begin{eqnarray}
Y_{ KK } & \sim & Y_5 \sqrt{k}
\end{eqnarray}
which 
(along with the above definition of $f$'s) explains
Eq. (\ref{y4d}). Similarly, the coupling of Higgs to
one zero-mode and one KK fermion is given by
\begin{eqnarray}
Y_{ mixed } ( c ) & \sim & 
Y_5 \sqrt{k} f(c)
\end{eqnarray} 
where
$c$ is that of the zero-mode fermion.
Finally,
the $c$-dependent part (which is the one relevant
for flavor-violation) of the coupling of two zero-mode fermions
to gauge KK mode is given by
\begin{eqnarray}
g^{ KK }_4 (c) & \sim & g_{ SM } \sqrt{ k \pi R } f(c)^2,
\label{gKK}
\end{eqnarray}
where we have used matching of $4D$ and $5D$ gauge couplings
at the tree-level and without any brane-localized kinetic terms 
for gauge fields.
We can use the above forumlae to estimate
charged lepton flavor violation in this framework which is of two types:
tree-level and loop processes which we now review in turn.

\subsection{Tree-level}
\label{lfvtree}

\begin{figure}
\centering
\includegraphics[scale=0.675]{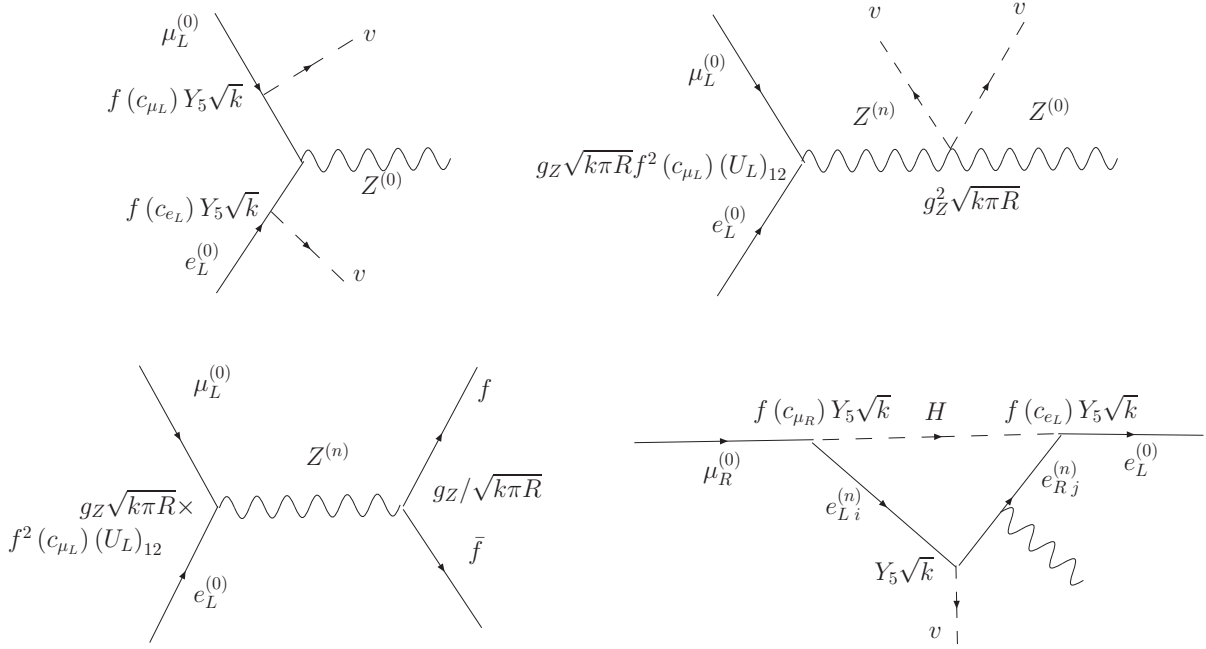}
\caption{\it Flavor violating couplings to $Z$ generated by zero-KK mode fermion mixing
(top left-hand side) and by zero-KK mode gauge mixing
(top right-hand side), $\Delta F = 1$
$4$-fermion
operators generated by exchange of gauge KK modes (with{\em out}
mixing with the zero-mode, bottom left-hand side) and dipole
operators generated by Higgs-KK fermion loops
(bottom right-hand side).}
\label{diagram}
\end{figure}

The tree-level flavor-violation occurs
dominantly
via $Z$ exchange with the
following flavor-violating $Z$ couplings to leptons (we focus on
$\mu$ and $e$ in this paper, but the formulae can be easily
generalized to the case of $\tau$'s):
\begin{eqnarray}
\delta g^Z_{ \mu_L e_L }
& \sim & \Big[ \frac{ M_Z^2 }{ M_{ KK }^2 } \times k \pi R
%
%r_c 
%
+ \frac{ \left( Y_5 \sqrt{k} \times v \right)^2 }{ M_{ KK }^2 } \Big]
\Big[ f \left( c_{ \mu_L } \right) \Big]^2\left( U_L \right)_{ 12 } 
\label{tree}
\end{eqnarray}
where 
1st term originates from mixing between zero
and KK gauge modes and 2nd term from fermion zero-KK mode 
mixing, both effects being induced by the Higgs vev: see Fig. 
\ref{diagram}.\footnote{We assume small brane-localized kinetic terms 
for $5D$ fields so that
the KK fermion and KK gauge masses are (almost) the same.}
Finally, $\left( U_L \right)_{ 1 2 }$
denotes the mixing angle of the
transformation from weak to mass basis for the charged leptons.

%%%%%%%%%%%%%%%%%%%%%%%%%%%%%%%%%%%%%%%%%%%%%%%%%%%%%%%%%%%%%%%

%%%%%%%%%%%%%%%%%%%%%%%%%%%%%%%%%%%%%%%%%%%%%%%%%%%%%%%%%%%%%%%

In particular, the assumption of a structureless or anarchic $Y_5$ 
(which we will make throughout this paper) implies that the mixing angles
between charged leptons are 
given by ratio of profiles at the TeV brane (i.e., $f$'s)
\begin{eqnarray}
\left( U_L \right)_{ i j } & \sim & \frac{ f \left( c_{ e_{ L \; i } } \right) }{ 
f \left( c_{ e_{ L \;  j } } \right) } \; 
\hbox{for} \; i < j
\label{mixing}
\end{eqnarray}
Similar formulae apply for $\delta g^Z_{ \mu_R e_R }$
and RH charged lepton mixing.

In the special case of
LH and RH charged lepton profiles being similar, i.e., 
hierarchies in charged lepton masses being explained {\em equally} by
ratios of RH and LH profiles at the TeV brane, 
we find that both mixing angles are small
and given by $\left( U_{ L, \; R } \right)_{ 12 } 
\sim \sqrt{ m_e / m_{ \mu } }$ (based on Eqs. (\ref{y4d}) and
(\ref{mixing})). 
We then obtain
\begin{eqnarray}
\delta g^Z_{ \mu_L e_L }, \; \delta g^Z_{ \mu_R e_R } & \sim & 
\Big[ \left( \frac{ M_Z^2 }{ M_{ KK }^2 } \times k \pi R
%
%r_c 
%
\right) 
\frac{ Y_{ 4 \; \mu } }{ Y_5 \sqrt{k} } + 
\frac{ Y_{ 4  \; \mu } Y_5 \sqrt{k} v^2 }{ M_{ KK }^2 } 
\Big] \sqrt{ \frac{ m_e }{ m_{ \mu } } }
\label{treeequal}
\end{eqnarray}
Note that there are 
flavor-{\em preserving} shifts in couplings to $Z$ which are
given by similar formulae (except there are no mixing angles
involved here).

\subsubsection{Direct KK $Z$ exchange}
\label{secdirectKK}

The
coefficient of the $4$-fermion operator $\sim \overline{ \mu_L } \gamma^{ \mu  } 
e_L \bar{f} \gamma_{ \mu } f$
(where $f =$ quark, lepton) generated by direct
exchange of KK $Z$ (i.e., without mixing with zero-mode
$Z$) is given by (see Fig. \ref{diagram})
\begin{eqnarray}
{\cal A}^{ \hbox{\small KK} \; Z } 
\left( \mu_L
\rightarrow e_L f \bar{f}
\right) 
& \sim &
\frac{ g_Z^2 }{ M_{ KK }^2 } 
\Big[ f \left( c_{ \mu_L } \right) \Big]^2\left( U_L \right)_{ 12 } 
\label{directKK}
\end{eqnarray}
where we have used the result that the 
flavor-{\em preserving} coupling of KK $Z$
is $\sim g_{ SM } / \sqrt{ k \pi R }$
and similarly for $\mu_R \rightarrow
e_R f \bar{f}$\footnote{KK photon will also induce similar effects.
And, in the models with
extended EW gauge symmetry, there is an addition neutral gauge
boson tower (denoted by $Z^{ \prime }$), i.e.,
the
combination of the
$U(1)$ subgroup of $SU(2)_R$ and
$U(1)_X$ which is orthogonal
to the hypercharge gauge symmetry, $U(1)_Y$. However,
flavor-{\em preserving} couplings of $Z^{ \prime }$ to light SM fermions
which are localized near the Planck brane are suppressed
compared to the coupling to KK $Z$ -- 
roughly the former couplings are of size 
given by $4D$ Yukawa couplings.}.

Comparing this effect to the one from $Z$ exchange (based on
Eq. \ref{tree}), we see
that the direct KK $Z$ exchange is 
suppressed by $k \pi R \sim \log \left( M_{ Pl } / \hbox{TeV} \right)$.
However, as mentioned earlier, 
we will invoke custodial symmetry to protect flavor violation
from $Z$ couplings, whereas direct KK $Z$ exchange
is not suppressed by this mechanism and thus
might become relevant in these cases.

\subsection{Loop}
\label{lfvloop}

The 
coefficient of dipole operator: $e \; F_{ \mu \nu } 
\overline{ \mu_{ L, \; R } }
\sigma^{ \mu \nu } e_{ R, \; L }$ induced by loops of KK fermions
and Higgs (including longitudinal $W/Z$), as in Fig. \ref{diagram}\footnote{It 
turns out that
the loops with KK $W/Z$ or transverse SM $W/Z$ and KK fermions
are approximately aligned with $4D$ Yukawa and hence do not
contribute to $\mu \rightarrow e \gamma$ \cite{Agashe:2004cp, Agashe:2006iy}.},
is given by 
\begin{eqnarray}
{\cal A} \left( \mu_R \rightarrow e_L \gamma \right)
& \sim & \frac{ \left( Y_5 \sqrt{k} \right)^2 }{ 16 \pi^2 }
\frac{ m_{ \mu } }{ M_{ KK }^2 } \left( U_L \right)_{ 12 } 
\label{loop}
\end{eqnarray}
and similarly for $\mu_L \rightarrow e_R \gamma$.

Again, 
in the case of LH and RH being similar, we find
\begin{eqnarray}
{\cal A} \left( \mu_R \rightarrow e_L \gamma \right), \; {\cal A} \left( \mu_L \rightarrow e_R \gamma \right)
& \sim & \frac{ \left( Y_5 \sqrt{k} \right)^2 }{ 16 \pi^2 }
\frac{ m_{ \mu } }{ M_{ KK }^2 } \sqrt{ \frac{ m_e }{ m_{ \mu } } }
\label{loopequal}
\end{eqnarray}

Note that there is some tension between tree-level and loop
processes from the size of $Y_5$ in the sense that the former (1st term in Eq.
(\ref{treeequal})) is enhanced for small $Y_5$ while the latter (Eq.
(\ref{loopequal})) is
suppressed in this limit.
Without considerations of neutrino data
(in particular, not taking into account the large charged current mixing
which is a combination of LH charged lepton and neutrino
mixing angles), we can 
assume LH and RH charged lepton profiles are similar, i.e.,
both sets of profiles are hierarchical at the TeV brane 
and mixing angles are small
as in Eqs. (\ref{treeequal}) and (\ref{loopequal}). This is the case studied in reference
\cite{Agashe:2006iy} with the result that the least 
constrained scenario (i.e., lowest KK scale) is with
$Y_5 \sqrt{k} \sim O( 
1
%
%\hbox{ a few } 
%
)$
which still requires 
%
%several (specifically, 
%
$\sim O(5)$\footnote{Note that this is the limit on KK scale obtained
by considering only {\em one} term at a time in the flavor-violating
amplitude (from among several uncorrelated terms of similar size),
whereas some of the limits quoted in reference \cite{Agashe:2006iy}
were based on the {\em combined} effect of all terms in this amplitude 
(for a certain choice of
relative phases between the various terms).} TeV gauge KK mass scale in order to be consistent
with charged lepton flavor violation data.
It turns out that the 
flavor-{\em preserving} shifts in $Z$ couplings to leptons are then quite safe.

\subsection{Enhanced effects due to fitting neutrino data}
\label{enhance}

Having estimated that
%
%several 
%
$\sim O(5)$
TeV gauge KK mass scale can be consistent with
charged lepton flavor violation with{\em out} considering neutrino masses, we
next discuss how incorporating neutrino
data affects these estimates.
It is
usually assumed that LH profile (at the TeV brane)
%
%entering
%
governing charged lepton 
mass is the same as that for 
neutrino mass (for each generation) 
because
LH lepton zero-mode
originates from a single $5D$ multiplet, i.e., 
$f \left( c_{ e_{ L \; i } } \right) = f \left(  c_{ \nu_{ L \; i } } \right) 
\equiv f \left(  c_{ L \; i } \right)$.
Clearly, along with the assumption of a
%
%structureless
%
anarchic $Y_5$, the mixing angles (appearing in the
bi-unitary transformation to go from
weak to mass basis) for LH charged leptons and neutrinos
are then of the same order (but not exactly the same) in these 
minimal models.
The reason for this feature is that the mixing angles are
dictated by 
%
%hierarchies in/
%
the ratios of
profiles of the three $L$ zero-modes near the TeV brane:
see Eq. (\ref{mixing}).
In turn, the 
neutrino oscillation data (i.e., large mixing in leptonic charged currents
which is 
a
combination of LH charged lepton and neutrino mixing) then requires
this LH lepton mixing angle to be large. 

Thus we make the following
change compared to the case without neutrino masses considered
in reference \cite{Agashe:2006iy}:
$\left( U_L \right)_{ 12 } \sim \sqrt{ m_ e / m_{ \mu } } \rightarrow
\sim O(1)$, which must result from no hierarchies in LH lepton profiles 
near the TeV brane, i.e., $f \left( c_{ L \; 1 } \right) \sim f \left( 
c_{ L \; 2 } \right)
\sim f \left( c_{ L \; 3 } \right)$.\footnote{We assume these
$f$'s are {\em not} exactly equal since that would require
a tuning of $c$'s.}
Thus, once we include neutrino data, it seems that 
LH and RH profiles cannot be
chosen to be similar for charged leptons. In turn, no hierarchies in
LH charged lepton profiles at the TeV brane implies that the 
hierarchies in charged lepton masses are
then explained {\em entirely} by hierarchies in RH charged lepton
profiles at the TeV brane, resulting in RH charged lepton mixing actually 
being smaller than in the case assumed in 
reference \cite{Agashe:2006iy}:
%
%sections \ref{lfvtree} 
%and \ref{lfvloop}:
%
$\left( U_R \right)_{ 12 } 
\sim \sqrt{ m_e / m_{ \mu } } \rightarrow 
m_e / m_{ \mu }$ (based on Eqs. (\ref{y4d}) and (\ref{mixing})). 

In short, with the above changes 
for mixing angles in the estimates for
charged lepton violation
from sections \ref{lfvtree} 
and \ref{lfvloop}, we find that the ``best'' case, i.e.,
with lowest KK mass scale, allowed
by charged lepton flavor violation
and taking into account the constraints from flavor-{\em preserving} shifts
$\delta g^Z_{ e_{ R \; i } e_{ R \; i } }, \;
\delta g^Z_{ e_{ L \; i }  e_{ L \; i } }, \; 
\delta g^Z_{ \nu_{ L \; i } \nu_{ L \; i } }
\stackrel{<}{\sim} \hbox{a few} \; 0.1 \%$
is the following: 
\begin{itemize}

\item
$M_{ KK } \sim O(10)$ TeV for
$Y \sqrt{k} \sim O(0.6)$, 
$f \left( c_{ \tau_R
%
%e \; 3 
%
} \right) \sim O(1)$ 
and
$f \left( c_{ L \; i } \right) \sim O(0.015)$ 

\end{itemize}
so that 
$c_{ L \; i } 
>
%
%\stackrel{>}{\sim} 
%
1/2$. Thus, the
LH lepton profiles are peaked near the Planck brane so that we do need to choose
$c_{ L \; i }$'s to be 
%
%almost 
%
%quite
%
%degenerate 
%
close to each other
in
order to achieve the (exponentially suppressed) profiles near the TeV brane 
being non-hierarchical: see Eq. (\ref{f})
(we will return to this
issue later).

In more detail (this discussion is an elaboration of
point (i) of section \ref{exacerbate} and will be useful later),
there are more than one ``count'' of enhancement of
charged lepton flavor violation 
once we include neutrino masses
relative to the case without neutrino masses:

\begin{itemize}
\item[I]
For $\mu_R \rightarrow e_L \gamma$, we have
enhancement from $\left( U_L \right)_{ 1 2 }
\sim \sqrt{ m_ e / m_{ \mu } } \rightarrow \sim O(1)$, although
$\mu_L \rightarrow e_R \gamma$ is suppressed compared to the
case without neutrino masses due to $\left( U_R \right)_{ 12 } 
\sim \sqrt{ m_e / m_{ \mu } } \rightarrow 
\sim m_e / m_{ \mu }$.
There is a similar enhancement and suppression for the two tree-level
contributions, i.e., $\delta g^Z_{ \mu_L e_L }$
and  $\delta g^Z_{ \mu_R e_R }$, respectively.
\item[II (A)]
Another count of enhancement (relative to case without neutrino masses)
for
$\delta g^Z_{ \mu_L e _L }$
%
%in addition to
%$\left( U_L \right)_{ 1 2 }
%\sim \sqrt{ m_ e / m_{ \mu } } \rightarrow O(1)$
%
comes due to three $f \left( c_{ L \; i } \right)$'s being similar, i.e., 
$f \left( c_{ L \; 2 } \right)$ in
Eq. (\ref{tree}) is clearly 
%
%influenced/
%
dictated by $m_{ \tau }$ 
(instead of
depending only on $m_{ \mu }$ earlier) since it is now (roughly) similar to 
$f \left( c_{ L \; 3 } \right)$ 
and hence can be larger.
\item[II (B)]
Moreover, we might try to choose smaller $Y_5$ 
in order to to suppress $\mu_R \rightarrow e_L \gamma$ (see Eq. (\ref{loop})), 
keeping 
several TeV KK mass scale
(in the light of point I above). Such
a smaller $Y_5$ implies that, in order to
obtain correct $m_{ \tau }$, $f \left( c_{ L \; 3 } \right)$ (and hence
$f \left( c_{ L \; 2 } \right)$ also), in turn, 
might have to be larger
than in the case without neutrino masses in reference \cite{Agashe:2006iy}.
\end{itemize}
Of course, we are free to choose RH and LH charged lepton profiles 
to be different
(unlike the case considered in reference \cite{Agashe:2006iy}), in particular,
we can increase $f \left( c_{ \tau_R
%
%e_R \; 3 
%
} \right)$ 
in order to make $f \left( c_{ L \; 3 } \right)$ smaller while keeping 
$m_{ \tau }$ fixed.
Hence $\delta g^Z_{ \mu_L e_L }$ can be smaller,
avoiding the
enhancements in point II (A) and (B) above. 
However,
a too large $f \left( c_{ \tau_R
%
%e_R \; 3 
%
} \right)$ is 
%
%limited
%
constrained by
$\delta g^Z_{ \tau_R \tau_R } \stackrel{<}{\sim} \hbox{a few } \; 0.1 \; \%$.
%
%For example, choose $Y_5 \sim$ to keep on edge for several TeV,
%but then $f_{ L \; 3 } \sim f_{ e \; 3 } \sim $ for $m_{ \tau }$, i.e., 
%both on edge of
%flavor-preserving shifts...$\delta g^Z_{ \mu_L e _L }$ requires...TeV
%
%lowest KK scale consistent with both processs from
%requiring KK limit from 2 processes same is then expected to be 
%in-between several and...
%TeV 
%
The best case is then obtained by 
choosing $f \left( c_{ \tau_R 
%
%e_R \; 3 
%
} \right)$ and $Y_5$\footnote{$f \left( c_{ L \; 3 }
\right)$ -- and hence
$f \left( c_{ L \; 1, \; 2 } \right)$ (up to
$\sim O(1)$ factor) -- is then fixed by
$m_{ \tau }$.} such that constraint on $M_{ KK }$ is the same from three
observables:
$\delta g^Z_{ \tau_R \tau_R }$, ${\cal A}
\left( \mu_R \rightarrow e_L \gamma \right)$
and $\delta^Z_{ \mu_L e_L }$ (it can be checked that the other processes -- both flavor-violating and
flavor-preserving -- are more easily satisfied and so are not the bottlenecks).
It is such an analysis
which shows that the lowest allowed KK scale is
$O(10)$ TeV (as mentioned above).

\section{Decoupling LH neutrino and charged lepton
mixing: new ``selection rules'' for Yukawa couplings}
\label{decoupling}

Clearly, the ``cornering'' involving the various observables discussed 
above can be simply avoided if
the LH profiles (at the TeV brane) which govern
the charged lepton and neutrino masses,
and hence the corresponding mixing angles, 
could actually be different.
A priori, it might seem difficult to achieve this
scenario
%
%possibility 
%
(due to LH charged lepton and neutrino
being $SU(2)_L$ partners) 
but remarkably it is possible as follows!
The central 
idea is that
the SM $SU(2)_L$ doublet LH lepton ($l^{ (0) } $) (for one generation) is actually 
a {\em combination} of zero-mode $SU(2)_L$ doublets from
two $5D$ multiplets with different profiles
such that the charged lepton masses originate from one component of this
zero-mode,
whereas obtaining neutrino masses requires using the other component. Although 
we focus on leptons here, a
similar argument can apply to quarks in order to obtain
parametrically different mixing angles for LH down-type vs. up-type
quarks.\footnote{In fact, two different $5D$ $SU(2)_L$
multiplets have already been used in references 
\cite{Contino:2006qr, Csaki:2008zd, Csaki:2008eh}
in order to obtain up and down-type quark masses,
%
%in the quark sector
%
but the implication
%
%application 
%
for decoupling the down-type quark mixing angles from the up-type  
was not specifically considered in these references.}

\subsection{General Case}

\begin{figure}
\centering
\includegraphics[scale=0.675]{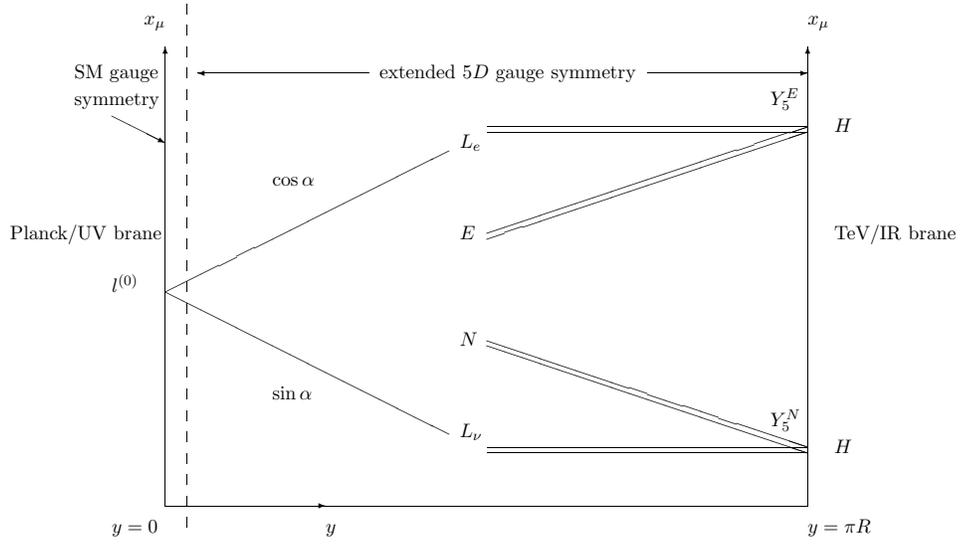}
\caption{\it The mechanism for decoupling LH charged lepton and neutrino
mixing angles. The single lines denote mixing of the
zero-modes from the bulk $SU(2)_L$ doublet 
multiplets, $L_{ e, \; \nu }$
on the Planck brane (which respects only the $4D$/SM
gauge symmetry) and double lines denote their couplings
(which respect the enlarged
%
%full 
%
$5D$ gauge symmetry)
to Higgs, $H$ and the bulk $SU(2)_L$ singlet multiplets,
$E$ and $N$.}
\label{trick}
\end{figure}

Let us see how to implement this idea in detail.
We will begin with a discussion of the general 
case which will enable us
to see how to apply it to other extra-dimensional models and also to the quark sector.
There are {\bf three} main ingredients of this idea
(which is summarized in Fig. \ref{trick}):
\begin{itemize}
\item[(1)]
Suppose the $5D$ gauge symmetry is {\em extended}
beyond the SM gauge symmetry and is 
%
%broken down/
%
reduced to
the SM gauge symmetry by boundary conditions at the Planck brane
(or equivalently by a large scalar vev on the Planck brane).
In other words, the gauge symmetry
of the $4D$ effective theory
(at the level of zero-modes) is only the SM symmetry, but it is a {\em subgroup}
of the $5D$ gauge symmetry.
\item[(2)]
Consider
two $5D$ fermion multiplets, $L_e$ and $L_{ \nu }$,
which
transform {\em differently} under the $5D$ gauge symmetry
(and hence cannot mix in the bulk/on the TeV brane). 
Moreover,
these two multiplets contain
zero-modes (to begin with: see later) -- denoted by $l^{ (0) }_{ e, \; \nu }$,
respectively --
%
%for components 
%
which
transform like LH leptons (i.e., identically) under the SM 
EW
%
%electroweak 
%
gauge symmetry.
Hence these two zero-modes can mix on the Planck brane ({\em only}) since 
the Planck brane 
respects only the SM (and not the full $5D$)
gauge symmetry.
%
%only one combination remains massless and is then
%identified with SM lepton, whereas the other one acquires a mass
%(with a Planck-brane localized fermion).
%

Specifically, one combination of the two zero-modes is given a (Planck-scale) mass
with a fermion localized on the Planck brane, $l^{ \prime }_{ R \; i}$
(effectively this
combination of the two $5D$ multiplets has Dirichlet
boundary condition on the Planck brane):
\begin{eqnarray}
{\cal L}_{ \hbox{UV brane} } & \ni & \overline{ l^{ \prime }_{ R \; i} } \left( \sin \alpha_i 
l^{ (0) }_{ e \; i }- \cos \alpha_i 
l^{ (0) }_{ \nu \; i } \right)
\label{massive}
\end{eqnarray}
The 
orthogonal combination of the two zero-modes is left over
as the only massless mode and is then identified with the SM LH
lepton:
\begin{eqnarray}
l^{ (0) }_i
& = & 
\cos \alpha_i l^{ (0) }_{ e \; i } + \sin \alpha_i 
l^{ (0) }_{ \nu \; i }
\label{massless}
\end{eqnarray}
For simplicity, we neglect flavor mixing in Eqs. 
(\ref{massive}) and (\ref{massless}).\footnote{In any case, we 
can show that such effects
not significant.
For the 
mixing of $l^{ (0) }_{ \nu }$ components of different generations
on the Planck brane,
this conclusion is due to either to
custodial protection for the resulting
flavor-violating couplings to $Z$ or to the choice $f \left( 
c_{ L_{ \nu } } \right) \ll 1$
(see discussion in section \ref{tuning}). Therefore, we are left with the
(mass) mixing of $l^{ (0) }_e$ components which can be shown
to be equivalent to mixing via Planck brane localized {\em kinetic} terms.
Such mixing appears even in the minimal models (and even without
consideration of neutrino masses)
where LH lepton arises from a single $5D$ field.
%
%it can be neglected 
%if we assume that Planck brane brane kinetic terms small
%
Flavor violation due to such kinetic terms (even if they
are $\sim O(1)$, i.e., comparable to
bulk contributions)
can also be shown to be
small.}
The gauge couplings of the SM fermion to leading order (i.e., 
couplings to the gauge zero-mode) are obviously not
affected by such a combination of the fermion zero-modes.
\item[(3)]
Moreover, the 
%
%transformations
%
representations of the
RH charged leptons and neutrinos
under the $5D$ gauge symmetry are {\em chosen} to be such that their 
couplings to Higgs 
-- localized near the TeV brane -- 
must
involve the two different components of the $l^{ (0) }$. 
The reason for these new ``selection rules''
is that the Higgs couplings (in general, all
bulk and TeV brane interactions) respect the full $5D$ gauge symmetry
(which is, again, larger than the SM one). 
Therefore, 
charged lepton and neutrino masses depend on the 
different profiles of the two components of $l^{ (0) }$, 
giving different LH
mixing angles.

\end{itemize}

Note that it is the enlarged $5D$ {\em symmetry}
which forces this ``decoupling''
of LH profiles
involved in the charged lepton masses
from those involved in the neutrino masses -- 
obviously the SM/$4D$ symmetry would allow
charged lepton and neutrino masses to proceed via the {\em same} component
of the $l^{ (0) }$.
Also, the Higgs couplings must satisfy
the larger $5D$ gauge symmetry even in 
the {\em minimal} case where we require (for simplicity) that the SM LH
lepton originates
as zero-mode of a {\em single} $5D$ field -- it is just that in this case
these selection rules then get 
translated into 
%
%particular
%
specific representations for the
RH charged leptons and neutrinos under the $5D$
gauge symmetry.

%%%%%%%%%%%%%%%%%%%%%%%%%%%%%%%%%%%%%%%%%%%%%%%%%%%%%%%%%%%%%%%

%%%%%%%%%%%%%%%%%%%%%%%%%%%%%%%%%%%%%%%%%%%%%%%%%%%%%%%%%%%%%%%

\subsection{Examples with extended electroweak 
symmetry: $SU(2)_L \times SU(2)_R \times
U(1)_X$}

Specifically, consider  
$SU(2)_L \times SU(2)_R \times U(1)_X$ as the $5D$ EW
%
%electroweak 
%
gauge
symmetry with $U(1)_Y$, being a combination
of $U(1)_X$ and $U(1)$ subgroup of $SU(2)_R$, i.e., $Y = T_{ 3 R } +
X$. As already mentioned, such an extension is motivated by
satisfying EWPT, in particular, the 
constraint from the $T$ parameter.
(However, this idea can be generalized to other extended $5D$ gauge symmetries,
such as a grand-unified one.)
The SM 
Higgs transforms as $\left( \bf{2}, \bf{2} \right)_0$,
where 1st/2nd symbol in $(...)$ is the representation under
$SU(2)_{ L, R }$ symmetry and the subscript denotes
the charge under $U(1)_X$.

The 
two different $5D$ multiplets which will constitute the
$l^{ (0) }$ transform as
$L_e: \left( \bf{2}, \bf{ r_{ L_e } } \right)_{ X_E}$
and $L_{ \nu }: \left( \bf{2}, \bf{ r_{ L_{ \nu } } } \right)_{ X_{ 
N
%
%\nu 
%
} }$,
respectively.
Note that, in general, $\bf{ r_{ L_{ e, \; \nu } } } \neq \bf{1}$ so that each 
multiplet can contain
{\em more than one} $SU(2)_L$ doublet.
We must choose the various charges such that 
$Y = T_{ 3 R } + X = -1/2$ (i.e., the $Y$ for the SM LH lepton)
for one $SU(2)_L$ doublet contained in each of the two multiplets,
$L_e$ and $L_{ \nu }$. Moreover, 
we choose only these two components of
the two multiplets to have zero-modes, i.e., to begin with, we
choose Neumann boundary condition on both branes only for the corresponding
$5D$ fields. Thus these two zero-modes correspond to the $l^{ (0) }_{ e, \; \nu }$
in the general discussion above.
Any extra ``would-be'' $SU(2)_L$ doublet zero-modes from the
rest of the multiplets can be projected out using Dirichlet boundary condition
for the corresponding $5D$ fields on the Planck brane.
Next, these two zero-modes
mix as in Eqs. (\ref{massive}) and (\ref{massless})
since on the Planck brane only $U(1)_Y$ is preserved so that finally we are
left with only one $SU(2)_L$ massless doublet (per generation).
Equivalently, ultimately only the 
combination of the two $5D$ multiplets $L_e$
and $L_{ \nu }$ in Eq. (\ref{massless}) 
has Neumann boundary condition
on the Planck brane.

The SM $e_R$ can arise as zero-mode of a {\em single}
$5D$ multiplet (denoted by $E$) and $\nu_R$ as zero-mode of a different 
(single) $5D$ multiplet (denoted by $N$).\footnote{Clearly we cannot do such a splitting 
for $e_L$ and
$\nu_L$ due to $SU(2)_L$ symmetry being preserved
at the zero-mode/$4D$ level (of course before Higgs acquires a vev).}
As discussed above, we
must choose 
representations for the RH charged lepton and neutrino
$5D$ multiplets under the bulk EW
gauge symmetry, $\left( \bf{1}, \bf{ r_E } \right)_{ X_E }$
and
$\left( \bf{1}, \bf{ r _{ 
%
%\nu 
%
N
} } \right)_{ X_{ 
%
%\nu 
%
N
} }$\footnote{As
for the case of $L_{ e, \; \nu }$, in general, we have $r_{ E,  \; 
%
%\nu 
%
N}
\neq \bf{1}$ so that there are extra components 
(other than the SM $e_R$ and $\nu_R$) in the $5D$ $E$ and
$
%
%\nu
%
N
$ multiplets. In fact, it is possible that
the $5D$ multiplet $E$ has a component with quantum numbers
of $\nu_R$ and vice versa. ``Unwanted'' zero-modes for such
components will have to be projected out 
using Dirichlet boundary condition on the Planck brane.
Again, only SM gauge symmetry is preserved on the Planck brane
so that such ``splitting'' of multiplets can be realized.} (respectively), 
such that charged lepton and neutrino masses must
proceed via the $l^{ (0) }_e$
and $l^{ (0) }_{ \nu }$ components of $l^{( 0) }$.
%
%zero-mode: 
%
Schematically, we desire
\begin{eqnarray}
\hbox{general case}: \; Y_E: \; \overline{ \left( \bf{2}, \bf{ r_{ L_e } } \right)_{ X_E } } 
\left( \bf{1}, \bf{ r_E } \right)_{ X_E } ( \bf{2}, \bf{2} )_ 0 & 
Y_{ N
%
%\nu 
%
}: \; \overline{ \left( \bf{2}, \bf{ r_{ L_{ \nu } } } \right)_{ X_{ 
%
%\nu 
%
N
} } } 
\left( \bf{1}, \bf{ r _{ N
%
%\nu 
%
} } \right)_{ X_{ N
%
%\nu 
%
} }
( \bf{2}, \bf{2} )_0 \end{eqnarray}
Thus, we require
\begin{eqnarray}
\bf{ r_{ L_e } } \times \bf{ r_E } \ni \bf{2} & \bf{ r_{ L_{ \nu } } } 
\times \bf{ r_{ N
%
%\nu 
%
} } \ni \bf{2}
\end{eqnarray}
so that
$E$ can couple to
$L_e$ 
%
%component 
%
and Higgs (and similarly for
$N
%
%nu
%
$), 
but 
\begin{eqnarray}
X_E \neq X_{ 
%
%\nu 
%
N
} \; \hbox{{\em or}} &
\bf{ r_{ L_e } } \times \bf{ r_{ 
N
%
%\nu 
%
} } \ni \! \! \! \! \not \; \; 
\bf{2}, \; \bf{ r_{ L_{ \nu } } } \times \bf{ r_E } 
\ni \! \! \! \! \not \; \; \bf{2}
\label{condition}
\end{eqnarray}
so that $E$ cannot couple to
$L_{ \nu }$ 
%
%component 
%
and Higgs (and similarly for
$N
%
%nu
%
$). Note that these Higgs couplings must preserve the enlarged,
i.e., $SU(2)_L \times SU(2)_R \times U(1)_X$, bulk
gauge symmetry,
and not just the SM symmetry. Hence, if Eq. (\ref{condition})
is satisfied, then charged lepton mass cannot proceed via the
$l^{ (0) }_{ \nu }$
component of $l^{ (0) }$ and vice versa.
Clearly, then the hierarchies in charged lepton and neutrino masses
are set by the hierarchies in profiles (at the TeV brane) 
of ($l^{ (0) }_e$, $e_R^{ (0) }$) and
($l^{ (0) }_{ \nu }$, $\nu_R^{ (0) }$), respectively.
In particular, 
large (small) LH mixing desired for
charged leptons (neutrinos) is achieved via small (large) hierarchies in
the profiles at the TeV brane 
of the $l^{ (0) }_e$ ($l^{ (0) }_{ \nu }$) components of the SM LH lepton.

\subsubsection{$X = 1/2 \; ( B - L )$}

We will now discuss 
%
%more specific
%
explicit examples. Begin with the
%
%canonical
%
minimal
%
%simplest 
%
case: $X = 1/2 \; ( B - L )$
and only one $5D$ LH lepton multiplet,
$L : \; (\bf{2}, \bf{1})_{ -1/2 }$
(of course this case will have the constraints discussed in
section \ref{enhance}).
The 
RH charged leptons and neutrinos
can be obtained from different $( \bf{1}, \bf{2} )_{ -1/2 }$ 
$5D$ multiplets
(labeled
with superscripts $e$ and $\nu$ below)
with the extra states in each multiplet 
having no zero-modes due to Dirichlet boundary condition
on the Planck brane.
%
%Higgs: $( \bf{2}, \bf{2} )_0$
%
Charged lepton and Dirac neutrino masses then both arise from
\begin{eqnarray}
\hbox{Case (0)}: & Y_E \; \hbox{and} \; Y_{ 
%
%\nu 
%
N
}: 
\; \overline{ ( \bf{2}, \bf{1} )_{ -1/2 } } ( \bf{1}, \bf{2})^{ 
%E
%
e, \; 
\nu 
%
%N
%
}_{ -1/2 }
( \bf{2}, \bf{2} )_ 0
\end{eqnarray}
giving similar (and hence large) mixing angle for LH charged leptons and
neutrinos and resulting in the KK mass limit $\sim O(10)$ TeV.
Recall that this choice of parameters additionally requires tuning
of $c_{ L \; i }$'s 
in order to obtain large mixings (as mentioned
earlier).

Note, however, that, even with the choice $X = 1/2 \; ( B - L )$,
all that we require is $T_{ 3 R }
= 0, -1/2, +1/2$ for $L$, $E$ and $N
%
%\nu
%
$, respectively, i.e.,
we are not forced to choose $L$ to be singlet of
$SU(2)_R$ or $
%
%e
%
E$, $N
%
%\nu
%
$ to be doublets of $SU(2)_R$
-- it suffices to choose integer and half-integer
spin representations of $SU(2)_R$, respectively, for
them. Thus we could instead choose
$L_e: ( \bf{2}, \bf{1} )_{ -1/2 }$, $L_{ \nu }: ( \bf{2}, \bf{5} )_{ -1/2 }$,
$E: ( \bf{1}, \bf{2} )_{ -1/2 }$ and $N:
%
%\nu
%
( \bf{1}, \bf{4} )_{ -1/2 }$ 
in order to 
satisfy the 2nd condition in Eq. (\ref{condition})
for different mixing angles (even though $X_E = X_{ 
N
%
%\nu 
%
}$
in this case). 
So, we have 
\begin{eqnarray}
\hbox{Case (1)}: & Y_E: \; \overline{ ( \bf{2}, \bf{1} )_{ -1/2 } } ( \bf{1}, \bf{2})_{ -1/2 } ( \bf{2}, \bf{2} )_ 0
& Y_{ N
%
%\nu 
%
}: \overline{ ( \bf{2}, \bf{5} )_{ -1/2 } } ( \bf{1}, \bf{4})_{ -1/2 } ( \bf{2}, \bf{2} )_0
\end{eqnarray}

\subsubsection{$X \neq 1/2 \; ( B - L )$}

In general, $X \neq 1/2 \; ( B - L )$ such that
even $X_E \neq X_{ 
%
%\nu 
%
N
}$ is possible. For example, 
$L_e: ( \bf{2}, \bf{1} )_{ -1/2 }$, $L_{ \nu }: ( \bf{2}, \bf{2} )_{ 0 }$,
$E: ( \bf{1}, \bf{2} )_{ -1/2 }$ and $N
%
%\nu
%
: ( \bf{1}, \bf{3} )_{ 0 }$ so that 
\begin{eqnarray}
\hbox{Case (2)}: & Y_E: \; \overline{ ( \bf{2}, \bf{1} )_{ -1/2 } } ( \bf{1}, \bf{2})_{ -1/2 } ( \bf{2}, \bf{2} )_ 0
& Y_{ N
%
%\nu 
%
}: \overline{ ( \bf{2}, \bf{2} )_{ 0 } } ( \bf{1}, \bf{3})_{ 0 } ( \bf{2}, \bf{2} )_ 0
\end{eqnarray}
As a final example of decoupling of LH charged and neutrino
mixing angles, we choose
$L_e: ( \bf{2}, \bf{2} )_{ - 1 }$, $L_{ \nu }: ( \bf{2}, \bf{2} )_{ 0 }$,
$E: ( \bf{1}, \bf{1} )_{ -1 }$ and $N
%
%\nu
%
: ( \bf{1}, \bf{1} )_{ 0 }$ so that 
\begin{eqnarray}
\hbox{Case (3)}: & Y_E: \; \overline{ ( \bf{2}, \bf{2} )_{ -1 } } ( \bf{1}, \bf{1})_{ -1 } ( \bf{2}, \bf{2} )_0
& Y_{ N
%
%\nu 
%
}: \overline{ ( \bf{2}, \bf{2} )_0 } ( \bf{1}, \bf{1})_0 ( \bf{2}, \bf{2} )_0
\end{eqnarray}
The motivation for the above two 
%
%choices
%
cases will be discussed later.

\section{Large neutrino mixing 
%
%with{\em out} 
%
with only mild
tuning}
\label{tuning}

The above idea of
decoupling large neutrino mixing
from LH charged lepton sector resolves only part of the problem
discussed in section \ref{enhance}, i.e., count (I) only, namely, the enhancing 
effect of mixing angles.
In particular, the EW structure of the charged lepton dipole operator
is 
similar to the masses
so that
the $l^{ (0) }_{ \nu }$ component of $l^{ (0) }$
does not enter the amplitude for $\mu \rightarrow e \gamma$.
Hence, the estimates for $\mu \rightarrow e \gamma$ are similar 
to the case without neutrino masses.

However, 
the flavor violation via tree-level $Z$ exchange
is still modified (relative to the case without neutrino masses)
as follows.
The point is that, although the
$l^{ (0) }_{ \nu }$ component of $l^{ (0) }$ does not determine the 
LH charged lepton mixing {\em angles}, it 
does 
%
%enter/affect 
%
contribute to 
the coupling of LH charged leptons to KK $Z$
which is (approximately) diagonal in generation space
in the weak basis for leptons\footnote{Here,
we are assuming that the off-diagonal couplings of leptons
(in this basis)
to {\em KK} $Z$ which are induced via brane-localized kinetic terms are small.
Similarly, such off-diagonal effects 
generated by zero-KK fermion mixing (see top left-hand side
of Fig. \ref{diagram},
with $Z^{ (0) }$ replaced by $Z^{ (n) }$)
are also suppressed,
assuming $Y_5
%
%^{N} 
%
\sqrt{k} \sim O(1)$. 
}. 
And, even though
we have $f \left( c_{ L_{ \nu \; 1 } } \right)
\sim f \left( c_{ L_{ \nu \; 2 } } \right) \sim f \left( c_{ L_{ \nu \; 3 } } 
\right)$, we are not assuming {\em strict} universality of these
$f$'s
(which would require tuning or a symmetry) so that couplings 
of SM LH leptons to KK $Z$ (see Eq. (\ref{gKK}))
induced by their $l^{ (0) }_{ \nu }$ components do differ
by $\sim O(1)$ factors.
In turn, via KK $Z$-zero-mode $Z$
mixing (see top right-hand side diagram in  Fig. \ref{diagram}),
these couplings to KK $Z$ result in 
{\em non}-universal  
shifts (relative to the $Z^{ (0) }$
coupling)
in the coupling of LH
charged leptons to SM $Z$. 
In the weak basis,
these shifts in the couplings to $Z$ are still diagonal in generation space.
Recall that
these non-universal shifts in coupling to $Z$ 
then get converted into flavor-{\em violating} coupling
to $Z$ up on going from weak to mass basis, as in
1st term of Eq. (\ref{tree})
(albeit with small mixing angle in this case). Therefore,
depending on the size of $f \left( c_{ L_{ \nu } } \right)$,
the $l^{ (0) }_{ \nu }$ component 
can still be important for tree-level charged lepton flavor violation
via $Z$ exchange.

We can then distinguish two cases (this discussion is an elaboration of
point (ii) of section \ref{exacerbate}).
If $c_{ L_{ \nu } } > 1/2$, then we have 
$f \left( c_{ L_{ \nu } } \right) \ll 1$, i.e.,
the $l_{ \nu }^{ (0) }$ are peaked near the Planck brane. 
In particular, we can choose 
$f \left( c_{ L_{ \nu } } \right) \stackrel{<}{\sim} f \left(
c_{ L_{ e \; 2 } } \right)$, where
the latter parameter determines $m_{ \mu }$, so that the effect of
$l^{ (0) }_{ \nu }$ component in tree-level 
charged lepton flavor violation (in zero-KK gauge mode mixing) 
is smaller than
that of the $L_e$ component\footnote{Similarly, the $l^{ ( 0 ) }_{ \nu }$
component of LH lepton also contributes to off-diagonal couplings
to SM $Z$ (already in the weak basis for leptons) via 
zero-KK fermion mode mixing. This effect can also be suppressed
by the choice of $c_{ L_{ \nu } } > 1/2$.}.
%
%the 
%charged lepton masses.
%
Then, the tree-level charged lepton flavor violation
is also same as in the case without neutrino masses, i.e., generically
KK mass limit is $O(5)$ TeV. 
%
%i.e., $O(5)$ TeV for cases (1)-(3) above (for further
%relaxation of constraint for case (3) see below)...
%
However, obtaining
non-hierarchical profiles near TeV brane for
the $l^{ (0) }_{ \nu }$ component
in order to generate large LH neutrino mixings then requires 
(almost) degenerate bulk masses, i.e.,
tuning, due to the profiles' exponential sensitivity to the bulk masses:
see Eq. (\ref{f}). Specifically, we need a splitting in $c$ of
$\sim 1 / \left( k \pi R \right) \sim 0.03$ if we require
(at most) a factor of
$\sim 3$ hierarchy in the profiles at the TeV brane for $c > 1/2$.

So, we consider instead $c_{ L_{ \nu } } \stackrel{<}{\sim} 1/2$
such that 
$f$'s have a milder (power-law instead of exponential) 
dependence on $c_L$ (see Eq. (\ref{f})), i.e., the $l_{ \nu }^{ (0) }$
have a flat/peaked near TeV brane profile.
Thus there is no need for 
any
%
%large 
%
tuning of bulk masses
in this case in order to obtain large LH neutrino mixing.
However, the case $c_{ L_{ \nu } } < 1/2$
is strongly constrained by 
EWPT
%
%electroweak precision tests
%
(independent of flavor-violation) even for
several TeV KK scale. Specifically, due to the
enhanced coupling of LH leptons to KK $Z$ (see Eq. (\ref{gKK})), the
flavor-{\em preserving} shift of the couplings
of leptons to $Z$\footnote{As discussed below, we can invoke custodial 
symmetries to suppress shifts in couplings of fermions to $Z$
in this case, but 
these can
only protect {\em either} (not both) LH charged lepton or neutrino couplings to $Z$
from being shifted.} and 
$4$-fermion operators induced by {\em direct} KK gauge 
exchange become too large\footnote{Explicitly, with dimensionless coefficient being
$O \left( g_Z^2 \right)$, such $4$-fermion operators
have to be suppressed by several TeV mass scale.}.

As a compromise, we are then led to considering $c_{ L_{ \nu } } \sim 1/2$ (i.e.,
close-to-flat profiles), but still 
not degenerate:
for example, based on Eq. (\ref{f}), we find that
\begin{itemize}

\item
$c = 0.525 \leftrightarrow 0.45$ gives only a factor of
$\sim 3$ (i.e., not larger) 
hierarchy in the profile at the TeV brane (see Eq. (\ref{f}))
which will still give large LH neutrino mixing. 

\end{itemize}
The splitting in bulk masses,
$\Delta c \sim 0.075$,
is still small,
but it is similar to the splitting of bulk masses in the quark 
sector (especially RH down-type) in {\em specific} models:
see, for example, reference \cite{Fitzpatrick:2007sa} for one possible   
fit of $c$'s to quark masses.
In this paper, we will accept 
%
%(or take the attitude that) 
%
this 
%
%(is) 
%
mild tuning.

With this choice, the 
flavor-{\em preserving} shifts in $Z$ couplings (see
Eq. (\ref{tree}) without mixing angle) are marginal, i.e., $\sim
\hbox{a few} \; 0.1 \%$, for several TeV KK scale
since $f \left( c_{ L_{ \nu \; i } }
\right)
\sim 1 / \sqrt{ \log \left( M_{ Pl } / \hbox{TeV} \right) }$  
(see Eq. (\ref{f})). 
And $4$-fermion operators induced by direct KK gauge exchange
are quite safe for several TeV KK
mass scale, using couplings in Eq. (\ref{gKK}). However, with this size of $f$'s, one problem is that 
the flavor-{\em violating} coupling
to $Z$, $\delta g^Z_{ \mu_L e_L }$ in Eq. (\ref{tree}), 
is still larger than in the case without 
neutrino masses (even with small mixing angle) -- in the latter
case, 
we get $f \left( c_{ L_2 } \right) \sim \sqrt{ m_{ \mu } / v }
\sim 1 / 40$
for the case of similar RH and LH charged lepton profiles and
$Y_5 \sqrt{k} \sim O(1)$.
Thus,
including neutrino masses is   
still dangerous on a count similar to (II A) in section \ref{enhance}
even though count (I) is avoided. Thus
we get the KK mass limit 
$> O(5)$ TeV from charged lepton
flavor violation. 
Similarly, for the minimal case (0) considered
earlier with the large LH charged lepton mixing angle, the
KK mass limit will be even larger than that mentioned before, i.e., 
$> O(10)$ TeV if we insist on 
no tuning, i.e., choose $c_L \stackrel{<}{\sim} 1/2$.

Note that we also need to generate non-hierarchical {\em mass splittings} for neutrinos
(in addition to mixing angles).
We can achieve
this goal by choosing
$c_{ N
%
%\nu 
%
} \stackrel{<}{\sim} 1/2$, i.e., mild or no tuning of
bulk masses for RH neutrinos giving non-hierarchical profiles
near the TeV brane. Combined with the non-hierarchical
LH neutrino profiles near the TeV brane (as above), the resulting 
Dirac neutrino masses will then be non-hierarchical, but too large
since both RH and LH profiles generating neutrino masses 
near TeV brane are larger than those of charged leptons.
However,
we can include (a Planck/GUT-scale) Majorana mass term
for RH neutrino on Planck brane 
and thereby use the see-saw mechanism \cite{Huber:2003sf, Perez:2008ee} 
to obtain very small 
neutrino masses (for other neutrino mass models in warped extra dimension,
see references \cite{Huber:2002gp, Carena:2009yt}).

\subsection{Custodial protection}
\label{custodial}

Next, we discuss how a custodial symmetry for the shift in the
coupling of fermions to $Z$ can relax
the above tension for the choice of close-to-flat profiles
for $l^{ (0) }_{ \nu }$ components of LH leptons.
In particular, we 
\begin{itemize}

\item
choose $L_{ \nu }: ( \bf{2}, \bf{2} )_0$ which implies
$T_{ 3 L } = T_{ 3 R }$ for this component
of the LH charged leptons. 

\end{itemize}
If we further choose
the $5D$ $SU(2)_{ L, \; R }$ couplings to be equal, 
then we realize the $P_{ LR }$ custodial symmetry \cite{Agashe:2006at} 
for this component of the LH charged leptons.
Such a symmetry
protects 
the couplings of SM $Z$ to leptons (in the {\em weak} basis)
from receiving a {\em non}-universal shift 
%
%(relative to the
%coupling to the zero-mode $Z$)
%
via zero-KK gauge mixing
(again on account of the $l^{ (0) }_{ \nu }$ component)
as follows
-- note that these shifts 
are flavor-{\em 
preserving}\footnote{due to 
the couplings of leptons 
to {\em KK} $Z$, $Z^{ \prime }$ being {\em diagonal} 
%
%(but 
%non-universal) 
%
in this basis, as mentioned earlier.}.
Recall that the couplings of LH charged leptons to KK $Z$ and similarly 
to (KK) $Z^{ \prime }$ 
induced by their 
$l^{ (0) }_{ \nu }$ components are {\em not} universal (although they have similar
size).
%
%
%Explicitly, 
%
However, 
this
symmetry enforces a cancellation between 
the non-universal contributions of KK $Z$ and KK $Z^{ \prime }$ 
in the mixing with $Z^{ (0) }$ so that the {\em net} shift in the
couplings of LH charged leptons (coming from their $l^{ (0) }_{ \nu }$
components)
to SM $Z$ is (approximately) universal, i.e., 
$\delta g^Z_{ e_{ L \; i } e_{ L \; i } }$ is $i$-independent.
Hence the resulting flavor-{\em violating} SM $Z$ coupling 
(after rotating from weak to mass basis for leptons) is suppressed as well
\footnote{It is clear that even in presence of off-diagonal
coupling of leptons
(in the weak basis) to {\em KK} $Z$, $Z^{ \prime }$ induced via 
brane-localized kinetic terms
(or zero-KK fermion mixing due to Higgs vev), 
this custodial protection for
flavor-violating lepton couplings to {\em SM} $Z$ still works since it 
is the result of a cancellation between
KK
$Z$ and $Z^{ \prime }$.
Similarly,
there is a cancellation between the contributions
of various KK fermions 
to the shift in coupling to SM
$Z$ from zero-KK fermion mode mixing
(see top left-hand side of Fig. \ref{diagram}) so that
this effect also enjoys custodial protection.}.
Note that we cannot simultaneously protect the (flavor-preserving)
$Z \overline{ \nu_L } \nu_L$ coupling from being shifted due to the 
$l^{ (0) }_{ \nu }$
component (due to $T_{ 3 L } = + 1/2 = - T_{ 3 R }$ for
$\nu_L$). However, as mentioned above,
in any case this effect is marginal (i.e., $\sim \hbox{a few} \; 0.1 \%$)
as long as $c_{ L_{ \nu \; i } } \sim 1/2$, i.e., $f \left( c_{ L_{ \nu \; i } } 
\right)
\sim 1 / \sqrt{ \log \left( M_{ Pl } / \hbox{TeV} \right) }$
in Eq. (\ref{tree}) (without mixing angle).\footnote{There is also a shift
in charged current lepton couplings (vs. those for quarks) due
to the mixing of KK and
zero-mode $W$ (especially
due to $l^{ (0) }_{ \nu }$ component of SM lepton), but again this effect 
is marginal. }
Also, note that the $\delta g^Z_{ \mu_L e_L }$ from
$l^{ (0) }_e$ component is {\em not} protected
(since $L_e$ must transform differently
under $SU(2)_R$ than $L_{ \nu }$, i.e.,
it must have $T_{ 3 R } \neq -1/2$,
in order to decouple LH
neutrino and charged lepton mixings), but anyway this effect is of similar 
size to the case
without neutrino masses and safe since we can choose $f \left(
c_{ L_{ e \; 2 } } \right) < 1 / \sqrt{ \log \left( M_{ Pl } / \hbox{TeV} \right) }$. 

We would like to 
%
%reiterate
%
emphasize that this mechanism to
suppress flavor-violating couplings to $Z$ does {\em not}
require the profiles at the TeV brane 
($f \left( c_{ L_{ \nu \; i } } \right)$'s)
and hence the couplings to KK $Z$ (see Eq. (\ref{gKK})) 
to be 
%
%degenerate
%
universal {\em at all}\footnote{again, we are assuming these
$f$'s are non-hierarchical in order to obtain
large neutrino mixing, but still differing by $\sim O(1)$ factors.},
but rather relies up on {\em cancellations} between
KK $Z$ and KK $Z^{ \prime }$
contributions (each of which
is {\em non}-universal) to the shifts in couplings of fermions to $Z$ . In this 
sense this mechanism to suppress
flavor-violating coupling to $Z$ is quite
distinct from the idea of $5D$ flavor symmetries which set $c$'s 
(and thus $f$'s)
to be degenerate.
Hence, with $5D$ flavor symmetries, couplings of leptons
to KK $Z$ 
(see Eq. (\ref{gKK})) and 
thus the
contribution of KK $Z$ 
to the shift in the coupling of
fermions to SM
$Z$ is {\em by itself} universal, giving
flavor-preserving $Z$ couplings after going from weak 
to mass
basis. Note that this result applies also to KK $Z^{ \prime }$
contributions, i.e., it is valid
{\em separately} for
KK $Z$ and KK $Z^{ \prime }$, unlike for the custodial 
symmetry case considered here.

{\bf Direct KK $Z$, $Z^{ \prime }$ exchange}: a detailed analysis
of this effect (including the effects
of $Z^{ \prime }$ which have not been calculated 
before\footnote{although the couplings of $Z^{\prime}$ to quarks,
relevant for $\mu$ to $e$ conversion in nuclei, are expected to
be negligible.})
is beyond the scope of this paper, but it suffices to
note that this
effect 
does not enjoy custodial protection
(unlike the effect of mixing of KK $Z$ with zero-mode $Z$).
Moreover, due to the $l^{ (0) }_{ \nu }$ component of LH lepton with
$c_{ L_{ \nu } } \sim 1/2$, i.e., $f \left( c_{ L_{ \nu } } \right) \sim 1 / 
\sqrt{ k \pi R }$, this effect can be
enhanced (for the LH leptons only) compared to
the case studied in reference \cite{Agashe:2006iy} without neutrino masses.
Specifically, 
with RH and LH charged lepton profiles
being similar, we get $f^2 \left( c_{ L \; 2 } \right) \sim m_{ \mu } / v
\sim 1 / 1700$ for $Y_5 \sqrt{k}
\sim O(1)$ in the latter model.
So, ratio of direct KK $Z$ exchange in the two models
is $\sim \sin^2 \alpha \times 1700 / ( k \pi R )$ based on Eq. 
(\ref{directKK}),
where 
$\sin \alpha$
is the admixture of $l_{ \nu }^{ (0) }$ in the SM
LH lepton as in Eq. (\ref{massless}).\footnote{Note 
that, 
as mentioned earlier,
we assume that the couplings of leptons (in weak basis) to 
{\em KK} $Z$, $Z^{ \prime }$ are (approximately) diagonal
(but non-universal)
in generation space.
%
%since, for simplicity
%we assume here
%effects of mixing brane kinetic terms are small and zero-KK fermion mixing.
%
So, the {\em off}-diagonal couplings to KK $Z$, $Z^{ \prime }$
giving flavor violation
arise only after rotating to mass basis:
the charged lepton mixing angle entering
this effect in the cases
we are considering here is the same as 
in the case without neutrino masses (i.e., this angle is small).}
As mentioned in section \ref{secdirectKK}, 
direct KK $Z$ exchange is suppressed
compared to $Z$ exchange in the model
without neutrino masses by $\sim k \pi R$
and latter is on the edge of data for $M_{ KK } \sim O(5)$ TeV.
Thus we see that direct KK $Z$ exchange in the
model under consideration here is
{\em marginal} if $\sin \alpha \sim 1$ and $M_{ KK } \sim O(5)$ TeV.

{\bf Choice of $L_e$ representations (responsible for charged
lepton masses)}: 
one possibility is 
$L_{ e }$: $(\bf{2}, \bf{1})_{ -1/2 }$ and $E$: $(\bf{1},
\bf{2})_{ -1/2 }$ as in {\bf case (2)}
above. The KK mass can then be as small as $O(5)$ TeV in case (2), 
even without 
any large tuning of $c_{ L_{ \nu } }$
in order to obtain large LH neutrino mixing angles.
If we allow tuning of $c_{ L_{ \nu } }$'s, 
we already saw in the beginning of section \ref{tuning}
that KK mass limit is same as that
without neutrino masses, i.e., $\sim O(5)$ TeV, as long as
we decouple LH charged lepton and neutrino mixings.
Recall that case (1) also has small LH charged lepton mixing angle
so that the KK mass limit can 
also be $\sim O(5)$ TeV, but this case does not have the custodial 
protection. So in case (1), we need to choose $c_{ L_{ \nu } }
> 1/2$ in order to suppress $\delta g^Z_{ \mu_L e_L }$
from $l^{ (0) }_{ \nu }$ component, 
implying that we need tuning for obtaining large
LH neutrino mixing angle.
A stronger limit on KK scale results if we instead choose $c_{ L_{ \nu } } \sim 1/2$ (i.e.,
no tuning) in case (1).

%
%i.e.,
%$M_{ KK } \stackrel{>}{\sim}$ TeV? (with large mixing angle,
%$M_{ KK } \stackrel{>}{\sim}$ TeV?)
%

\subsubsection{Best case scenario}
\label{best}

So far, we have been able to obtain a {\em similar} level of charged lepton
flavor violation (and hence
$\sim O(5)$ TeV KK scale) as in the case without neutrino masses discussed in 
reference \cite{Agashe:2006iy}.
In fact, we can obtain {\em more} safety (relative to the case
without neutrino masses in reference \cite{Agashe:2006iy}) using  
{\bf case (3)} above which has 
$T_{ 3 L } = T_{ 3 R } = 0$ for RH charged leptons.
Such a choice of quantum numbers results in 
custodial 
protection ($P_C$ symmetry \cite{Agashe:2006at}) for non-universal shifts in
couplings of $Z$ to RH charged leptons (in weak basis). As for the
$P_{ LR }$ custodial 
symmetry discussed before, there is 
is a cancellation between the non-universal
contributions of KK $Z$ and (KK) $Z^{ \prime }$ in the mixing
with zero-mode $Z$\footnote{again, 
a similar effect occurs for zero-KK mode fermion mixing.}, but
we do not need the $5D$ $SU(2)_{L, \; R}$ gauge couplings
to be equal in this case (unlike for $P_{ L R }$ symmetry).
The idea then is to 
\begin{itemize}

\item
increase {\em all} $f \left( c_E \right)$'s by, say, $2 \sqrt{2}$ 
compared to the case
in reference \cite{Agashe:2006iy} 
-- note that 
flavor-{\em violating} coupling to $Z$, i.e., $\delta g^Z_{ \mu_R
e_R }$ (after rotating from weak to mass
basis), is also protected by this custodial
symmetry
and 

\item
reduce {\em all} $f \left( c_{ L_{ e \; i } } 
\right)$ 
by
$\sim \sqrt{2}$
-- recall that $\delta g^Z_{ \mu_L e_L }$
resulting from $L_e$ component is not protected since
we have $T_{ 3 L } \neq T_{ 3 R }$
for this component of LH charged lepton. 

\end{itemize}
Hence, we can 
reduce $Y_5$ by $\sim 2$, keeping charged lepton masses fixed.
Then both the  
tree-level\footnote{Again, the LH contribution of Eq. (\ref{tree}) is 
suppressed only by
$\sim 2$, but there is no RH contribution due to custodial
protection, giving another reduction by factor of $2$.}
and loop amplitudes are reduced by $\sim 4$ (see Eqs. (\ref{tree})
and (\ref{loop}))
compared to the case in
reference \cite{Agashe:2006iy}. KK mass scale can then be smaller
by $\sim 2$, i.e., $O(2.5)$ TeV.\footnote{We 
chose hierarchies in
%
%ratios of 
%
LH charged lepton
profiles (at the TeV brane) from $l^{ (0) }_e$ component
to be similar to those of RH charged leptons, 
i.e., both RH and LH charged lepton $1-2$ mixing
angle $\sim \sqrt{ m_e / m_{ \mu } }$, as in reference \cite{Agashe:2006iy}.
It is easy to check that such a choice minimizes
the constraint from $\mu \rightarrow e \gamma$.}
If we keep increasing $f \left( c_E \right)$'s even more, 
then, eventually, RH charged lepton
flavor violation from {\em direct} KK $Z$ exchange
(which is {\em not} protected) becomes relevant. 
%
%but there is some room
%to play with 
%here.
%
Thus, we conclude that 
%
%a few (say, 
%
$\sim O(3)$ TeV KK scale can be 
%
%{\em easily} 
%
consistent in case (3)
with both charged lepton flavor violation and neutrino masses and 
with at most mild tuning of $c_{ L_{ \nu } }$'s in order to obtain large
neutrino mixing.\footnote{Based on the previous
discussion, it can be seen that a
{\em very} mild tuning of $\sin \alpha$ is required
to make the effect of direct KK $Z$ exchange, coming
from $l^{ ( 0 ) }_{ \nu }$
component of LH leptons, marginal
for $\sim O(3)$ TeV KK scale.}

\subsubsection{A case with custodial protection, but {\em no} decoupling of mixing angles}

%
%As an aside and 
%
In order 
to illustrate the independence of the above two 
mechanisms, namely, decoupling
of large neutrino mixing from charged lepton sector
(discussed in section \ref{decoupling}) and custodial symmetry
studied in this section, we consider a final case with
only one $5D$ multiplet for LH leptons,
but
choose $L: ( \bf{2}, \bf{2} )_{ 0 }$ for custodial protection
for LH charged leptons. 
With $E: ( \bf{1}, \bf{3} )_0$
and $N
%
%\nu
%
: ( \bf{1}, \bf{1} )_0$, we get 
\begin{eqnarray}
\hbox{Case (4)}: & Y_E: 
\; \overline{ ( \bf{2}, \bf{2} )_0 } ( \bf{1}, \bf{3})_0 
( \bf{2}, \bf{2} )_0
& Y_{ 
N
%
%\nu 
%
}: \overline{ ( \bf{2}, \bf{2} )_0 } ( \bf{1}, \bf{1})_0 ( \bf{2}, \bf{2} )_0
\end{eqnarray}
The 
large, i.e., $O(1)$, mixing both for LH charged leptons and neutrinos
due to the choice of one $5D$ $L$ multiplet implies that we must choose 
$Y_5$ smaller (i.e., $\sim O(1/4)$)
so that 
%
%several, i.e., 
%
$\sim O(5)$ TeV KK mass scale can be allowed
by $\mu \rightarrow e \gamma$ (see Eq. (\ref{loop})). 
In order to compensate the effect
of smaller $Y_5$ in $\tau$ Yukawa coupling, we can then increase
$f \left( c_{ E \; 3 } \right)$ such that we are on 
the edge of the $\delta g^Z_{ \tau_R \tau_R }$ constraint for
several TeV KK mass scale, i.e., $f \left( c_{ E \; 3 } \right) \sim
1 / \sqrt{ k \pi R }$
(recall that we
do {\em not} have custodial protection for RH charged lepton couplings to $Z$
in this case since $T_{ 3 R } = -1$). Even with
this extreme choice of $f \left( c_{ E \; 3 } \right)$, we still need
$f \left( c_{ L } \right) \sim 1 / \sqrt{ k \pi R }$, i.e., $c_L \sim 1/2$ 
in order to obtain $m_{ \tau }$.
Anyway, $c_L \sim 1/2$ is favored by the desire to 
obtain large LH mixing angles with no tuning.\footnote{Also, we cannot
choose $c_L < 1/2$ in spite of custodial protection
for LH charged lepton coupling to $Z$ since we do not simultaneously
have such protection for $\nu_L$ couplings.}
We can check that the 
tree-level 
$\delta g^Z_{ \mu e }$ is quite safe, due to custodial protection for LH
contribution and due to (very) small mixing angle
for RH contribution (even for such
large $f \left( c_{ E \; 3 } \right)$).

Based on the previous discussion, it is clear
that the flavor violation from direct KK $Z$ exchange
%
%$Z^{ \prime }$ 
%
in this case
violates the experimental constraint by $\sim O(10)$ due to mixing angle
being larger than in cases (2) and (3) discussed above.
A mild tuning $\sin^2 \alpha \sim
O(0.1)$ can make this effect marginal for
$M_{ KK } \sim O(5)$ TeV in case (4).
Note that we could also have invoked 
$\sin \alpha \ll 1$ in order to obtain a suppression
for $Z$ exchange (instead of using custodial protection), 
but clearly we would have needed
significant tuning in this case.
%
%ways out (can also be used in case 2, 3)
%
%-- $Z^{ \prime }$ for LH a bit of cancellation in charge, negligible for RH
%leptons and quarks ($\mu$-to-$e$
%conversion in nuclei
%dominant constraint)
%
%-- KK $Z$ cancellation for $L_{ \nu }$ component due to $\sim 1/2$
%
%-- only LL RR from KK $Z$, photon instead of 4 combinations for $Z$?
%

The various possibilities discussed in this paper are 
summarized in table \ref{choice}.
Clearly, other possibilities with low KK scale can
be constructed from {\em combinations}
of the cases presented in this table.
\begin{table}
\begin{center}
\caption{The representations ${\bf r_{ L_e }}$
and ${\bf r_{ L_{ \nu } }}$ under the
bulk gauge symmetry, $SU(2)_L \times SU(2)_R \times U(1)_X$,
for the two components of LH leptons, $l^{ (0) }_e$ and
$l^{ (0) }_{ \nu }$, respectively, discussed in the text.
The
``Y'' and ``N'' in 3rd column convey whether charged lepton mixing
angle is small or not. Similarly they convey whether the 
case has custodial symmetry
for the  $l^{ (0) }_{ \nu }$ component of LH lepton
with non-hierarchical profiles at the TeV brane (which
give neutrino masses with large mixing) or not (4th column)
and finally 
custodial symmetry for RH charged lepton multiplet (5th column).
The
last column shows lower limit on $M_{ KK }$ from charged
lepton flavor violation. 
For the cases
with small LH charged lepton mixing, we are
considering flavor-violating contributions from 
$l^{ (0) }_e$ component of LH lepton (which give charged lepton masses)
and 
%
%$e$ 
%
RH charged lepton multiplets
only, i.e., contribution from the non-hierarchical 
$l^{ (0) }_{ \nu }$ profiles is assumed to be negligible in
these cases. This assumption is justified
due to either (i) the choice of these profiles peaked near Planck brane,
%
%$c_{ L_{ \nu } } > 1/2$ 
%
which requires
tuning of bulk masses in order to obtain large neutrino mixing or 
(ii) presence of custodial protection (i.e., ``Y'' in 4th column)
for the case of close-to-flat profiles, which does not require tuning.}

\label{choice}

\begin{tabular}{|c|c|c|c|c|c|}
\hline 
$\bf{ r_{ L_e } }$ & 
$\bf{ r_{ L_{ \nu } } }$ &
small mixing angle? &
$L_{ \nu }$ custodial? &
RH custodial? & 
lower limit on $M_{ KK }$
\tabularnewline
\hline
$( \bf{2}, \bf{1} )_{ -1/2 }$ & 
$( \bf{2}, \bf{1} )_{ -1/2 }$  & 
N &
N &
N &
$\sim O(10)$ TeV
\tabularnewline
\hline
$( \bf{2}, \bf{1} )_{ -1/2 }$ &
$( \bf{2}, \bf{5} )_{ -1/2 }$ & 
Y &
N &
N &
$\sim O(5)$ TeV
\tabularnewline
\hline
$( \bf{2}, \bf{1} )_{ -1/2 }$ & 
$( \bf{2}, \bf{2} )_{ 0 }$ &
Y & 
Y &
N &
$\sim O(5)$ TeV
\tabularnewline
\hline
$( \bf{2}, \bf{2} )_{ -1 }$ & 
$( \bf{2}, \bf{2} )_{ 0 }$ & 
Y & 
Y &
Y &
%
%$\sim$ a few
%
$\sim O(3)$ TeV
\tabularnewline
\hline 
$( \bf{2}, \bf{2} )_{ 0 }$ & 
$( \bf{2}, \bf{2} )_{ 0 }$ & 
N &
Y &
N &
$\sim O(5)$ TeV
\tabularnewline
\hline

\end{tabular}

\end{center}

\end{table}

\section{Conclusions
and Outlook
%
%Discussions
%
}
\label{concl}

As we eagerly await the start of the LHC, where new physics at the TeV scale related to
Planck-weak hierarchy of the SM might be discovered, it is
interesting to study whether clues of {\em flavor} hierarchy 
of the SM could lie in this physics.
In this paper, we considered one such possibility, namely, the
framework of a warped 
extra dimension with the SM gauge and fermion
fields propagating in the bulk. The flavor hierarchy of
the SM can be accounted for in this framework using profiles
for the SM fermions in the bulk, but the flip side is the
resulting 
flavor violation from KK modes. Even though there is
an automatic GIM-type mechanism, the limit 
on KK mass scale from flavor violation in both the quark and charged lepton sector 
(with{\em out} considerations
of neutrino data) is still 
%
%several 
%
$\sim O(5)$ TeV.

%
%In particular
% 
Moreover, if we include the neutrino data, then
the charged lepton flavor violation tends to be 
enhanced by the large charged current mixing required to account
for neutrino oscillations. The point is that, in the minimal model, the 
mixings are similar for LH charged leptons and neutrinos, being dictated by 
LH profiles (at the TeV brane)
which are same for the two sectors. Hence, the limit on
gauge KK mass scale from charged lepton flavor violation
{\em when combined with neutrino data} is 
{\em larger than}
%
%several 
%
$\sim O(5)$ TeV, 
%
%highly suppressing 
%
making any signals at the LHC from 
direct production of gauge KK modes unlikely. 

In this paper, we presented new mechanisms which can suppress 
charged lepton flavor violation in this framework.
The central point is to use
less minimal representations
for leptons under the extended $5D$ gauge symmetry\footnote{such an
extension of $5D$ gauge symmetry is motivated for 
satisfying EWPT.}, 
allowing mixing angles to be (simultaneously) 
small and large in the LH charged lepton
and neutrino sectors, respectively. The trick is that the
LH lepton zero-mode is actually a combination of two zero-modes
with different profiles, 
one giving charged lepton masses and the other neutrino masses.
Furthermore, such representations can lead to custodial protection
for the shift in couplings of charged leptons to $Z$ (ala $Z b \bar{b}$)
and hence suppress charged lepton flavor violation
from tree-level exchange of $Z$.

The bottom line is that 
%
%a few 
%
$\sim O(3)$
TeV gauge KK mass scale might then allowed by charged lepton
flavor violation, including neutrino masses
and without any particular structure in the $5D$ flavor
parameters. However, 
charged lepton flavor violation is still
not ``super-safe'' (unlike in some models
with $5D$ flavor symmetries) so that the upcoming lepton flavor violation
experiments (MEG at PSI \cite{Signorelli:2003vw}, PRIME at 
JPARC \cite{PRIME} and the proposed mu2e
experiment at Fermilab \cite{mu2e}) 
should see a signal.
The situation is 
similar to reference \cite{Agashe:2006iy} without considerations of neutrino masses 
since the two issues
of charged lepton flavor violation and neutrino masses are now decoupled.
Also, with 
%
%a few 
%
$\sim O(3)$
TeV gauge KK scale, signals form direct production 
of these KK modes at the LHC are then still viable \cite{kkgluon}.

\subsection{Other applications}

We would like to emphasize that the 
mechanism discussed in this paper is quite general 
%
%in several ways
%
as we discuss below with several examples. 

{\bf Quark sector in warped extra dimensional framework}:
the mechanism for decoupling of mixing angles 
of the LH charged leptons and neutrinos can be applied
to quark $SU(2)_L$ doublets as well
for the warped extra-dimensional scenario. In particular, we can arrange for
LH down-type quark mixing to be {\em parametrically} smaller than
LH up-type quark mixing -- the latter would then have to entirely account for
the CKM mixing. Thus flavor-violating
%
%operators 
%
effects
involving LH down-type quarks 
% 
%$(V-A) \times (V-A )$ 
%
%from neutral gauge KK exchange
%
can be suppressed compared to the minimal models, where the 
LH down-type and up-type quark mixings are similar (just like
for LH charged leptons and neutrinos) and hence of CKM-size.

However, the dominant constraint on gauge KK scale
from flavor violation in the quark sector comes
from contributions to 
$\epsilon_K$ involving {\em both} LH and RH down-type quarks.
While LH down-type quark mixings can be suppressed
using the trick used for leptons here, it is easy to see that the
the RH down-type quark mixings are enhanced compared to minimal models\footnote{just like we found for charged leptons in section \ref{enhance}
that enhancement of LH mixing angles implies reduction
in RH mixing angles.} 
so that this mixed
contribution to $\epsilon_K$ is not affected.
However, the dominant contribution to $B_{ d, \; s }$ mixing
does come from operators with LH down-type quarks only and hence
it can be suppressed using the mechanism of decoupling
LH up and down-type quark mixing angles. Of course, the
constraint on KK mass scale from these systems is (generically) 
weaker than the one from $\epsilon_K$.
In short,
it seems difficult to {\em fully}
ameliorate the constraints from quark sector flavor violation using this mechanism.

{\bf Combining with other proposals within
warped extra dimensional framework}:
we have presented the new mechanisms for suppressing
charged lepton flavor violation in the warped extra dimensional
framework in a manner showing their 
independence from other ideas in the literature.
%
%stand on their own.
%
However, it is clear that these mechanisms
can actually be combined with other ideas.
For example, reference \cite{Agashe:2008fe} 
proposed obtaining (naturally) small and large mixing angles for LH charged
leptons and neutrinos, respectively, 
even with {\em minimal} choice of
representations of the bulk gauge symmetry,
in a framework where neutrinos are Dirac particles and with a bulk 
Higgs. In this framework, we choose non-degenerate $c_L > 1/2$, i.e., with
LH lepton profiles being
hierarchical near the TeV brane and non-hierarchical near
the Planck brane. The point is that we can then obtain
small charged lepton mixing angles (as usual) since charged lepton
masses are dominated by overlap with Higgs near the TeV brane, whereas
neutrino masses
can be dominated by overlap of profiles near the Planck brane thus
giving large neutrino
mixings.
Thus, even
without using the new $SU(2)_R$ representations
(i.e., the two mechanisms of this paper), the lepton flavor violation
constraints in this framework
for neutrino masses reduce to
the case without neutrino masses
studied in \cite{Agashe:2006iy}, i.e., the gauge KK mass limit is
$\sim O(5)$ TeV.
In addition, the mild tuning of $c_L$'s required in the models
considered here in order to obtain large neutrino mixings
is avoided in the idea of reference \cite{Agashe:2008fe} .

Interestingly, we can add the custodial protection for 
charged lepton couplings to $Z$ to the above idea.
Namely, we move 
either LH or RH charged lepton profile closer to the TeV brane (relative to the
choice of same RH and LH profiles), invoking
custodial symmetry 
to protect tree-level flavor violation via $Z$
exchange from this chirality. 
Simultaneously, we move the profile of the other 
chirality (which does not enjoy custodial protection)
away from
the TeV brane (thus reducing tree-level $\mu$ to $e$ conversions) 
in such a way as to
allow a reduction in $5D$ Yukawa coupling and thus suppressing,
in turn, loop-induced $\mu \rightarrow e \gamma$ as well.
We note that 
\begin{itemize}

\item
such a strategy (along
the lines discussed in section \ref{best}) 
can 
%
%rather easily 
%
allow us to lower the limit (from lepton flavor violation)
on the gauge KK scale
in this framework 
from $\sim O(5)$ TeV (which is the value without custodial
protection) down to $\stackrel{<}{\sim} O(3)$ TeV. 

\end{itemize}
Similarly, these
mechanisms can be suitably
combined with $5D$ flavor symmetries. We will leave these directions
for future work.

{\bf Beyond applications to the warped extra dimensional framework}:  
these mechanisms might enable
suppression of flavor violation (especially in
charged lepton sector) in 
other extra-dimensional models
which explain flavor hierarchy via profiles, as long as there
is an extended gauge symmetry to play the decoupling trick.
In particular, 
another framework where the origins of 
flavor 
%
%physics is 
%
%brought down 
%
leave their imprint on
%
%to
%
physics at the 
TeV scale (i.e., within
LHC reach) is the
recently proposed $5D$ flavorful SUSY \cite{Nomura:2007ap}.
This $5D$ set-up can be quite similar to that considered in this paper, namely
Higgs localized on one brane in an extra dimension with light fermion
profiles being peaked near the other end of the extra dimension -- the
smallness of the fermion profiles near the Higgs brane then account for
the lightness of these fermions.

More importantly, 
SUSY breaking can occur on the Higgs brane in this framework such that the 
non-universalities/mixing among 
squarks and sleptons are governed by the (s)fermion
profiles
near Higgs brane. Thus the structure of squark and slepton masses
is correlated with the SM Yukawa couplings,
possibly suppressing SUSY contributions
to flavor violation at least for the 1st/2nd generation. This
effect for $5D$ flavorful
SUSY is the analog of the GIM-like mechanism for KK 
contributions considered here -- of course, the
{\em actual} KK contributions could be much smaller 
for $5D$ flavorful SUSY due
to higher compactification scale with the resulting hierarchy 
between that scale and the weak scale being explained by SUSY.
However, charged lepton flavor violation in $5D$ flavorful SUSY could be 
enhanced due to large neutrino mixing just like discussed here.
It will be 
interesting to further study the mechanisms
for suppressing 
%
%charged lepton 
%
flavor violation 
discussed in this paper
in the context of $5D$ flavorful SUSY. 
%
%and for quark sector
%in the warped extra dimensional framework.
%

\section*{Acknowledgments}

This work is supported by 
NSF grant No. PHY-0652363.
The author would like to thank Roberto Contino, 
Takemichi Okui, Gilad Perez and Raman Sundrum
for discussions, Gilad Perez for comments on the
manuscript
and the Aspen Center for Physics for hospitality during part of this work.

%%%%%%%%%%
%%%%%%%%%%    References
%%%%%%%%%%

\end{document}